\documentclass[twocolumn,samsmath,amssymb,floatfix,unsortedaddress,aps]{revtex4-1}
\usepackage{amsmath}
\usepackage{fixmath}
\usepackage{latexsym,amsfonts,mathrsfs,color,graphicx}

\begin{document}
\title{Morphology of clusters of attractive dry and wet self-propelled spherical particle suspensions} 

\author{Francisco Alarc\'on}  \affiliation{Departament de F\'isica de la Mat\`eria Condensada, Universitat de Barcelona, C. Mart\'i Franqu\'es 1, 08028-Barcelona, Spain}
 \affiliation{University of Barcelona Institute of Complex Systems (UBICS), Universitat de Barcelona, Barcelona, Spain.}

\author{Chantal Valeriani} \affiliation{Departamento de Fisica Aplicada I, Facultad de Ciencias Fisica, Universidad Complutense de Madrid, 28040 Madrid, Spain.} 

\author{Ignacio Pagonabarraga} \affiliation{Departament de F\'isica de la Mat\`eria Condensada, Universitat de Barcelona, C. Mart\'i Franqu\'es 1, 08028-Barcelona, Spain}
 \affiliation{University of Barcelona Institute of Complex Systems (UBICS), Universitat de Barcelona, Barcelona, Spain.}
 


\begin{abstract}

In order to asses the effect of hydrodynamics in the  assembly of 
active attractive spheres, 
we simulate a semi-dilute suspension of attractive self-propelled spherical particles  in a quasi two dimensional 
geometry comparing the case with and without  hydrodynamics interactions. 
To start with, independently on the presence of hydrodynamics, we observe that depending on the ratio between attraction and propulsion, 
particles  either coarsen or aggregate forming finite-size clusters.  
Focusing on the clustering regime, we characterize two different 
 clusters parameters, i.e. their morphology and orientational order, and compare  
 the case when active particles behave either as pushers or pullers 
(always in the regime where inter-particles attractions competes with self-propulsion).
Studying cluster phases for squirmers with respect to those obtained for active Brownian disks (indicated as ABP),  
 we have shown that hydrodynamics alone can sustain a cluster phase of active swimmers (pullers), while ABP form cluster 
phases due to the competition between attraction and self propulsion. The structural properties of the cluster phases of squirmers
 and ABP are similar, although squirmers show  sensitivity to active stresses. Active Brownian disks  resemble weakly pusher squirmer 
 suspensions in terms of cluster size distribution, structure of the radius of gyration on cluster size and degree of cluster polarity.




\end{abstract}
\maketitle

\twocolumngrid

\section{Introduction}

Active matter is concerned with the study of systems composed of self-driven units and 
active particles,  able of converting energy into systematic movement \cite{schweitzer}.
Examples of active matter are found from micro to nano length scales, in living or nonliving systems, such as  cells, tissues and living organisms, animal groups, self-propelled colloids and artificial nanoswimmers \cite{lauga2009,aronson,dileoreview}.  One important feature of active matter is that its elements can develop coordinated behaviour, such as collective motion \cite{vicsek_rev}. 

Experiments in this field are  developing at a rapid pace \cite{frenchexp, sashi, beer,libchaber2015} and a new theoretical framework is 
needed to establish a ``universal'' behaviour among these internally driven systems. 
\textcolor{black}{With this goal in mind, a suspension of self-propelled   Brownian particles 
with  short-range repulsions, no hydrodynamic interactions (HI) and a spherical shape (thus 
no steric effects induced by an elongated shape) has been considered as a} \textcolor{black}{ characteristic,  reference system.} 


\textcolor{black}{
Given a suspension of \textcolor{black}{spherical} repulsive active Brownian particles (ABP), it has been shown  that one can derive an expression for the pressure
 \cite{Mallorypre2014,Bradyprl2014,Bradypre2015,Solonprl2015,Gompperviral2015}, even  though   
 the mechanical pressure  might not lead to an equation of state \cite{solonNature}.
When dealing with dense suspensions, both 
 numerical simulations and theory have   identified a motility 
induced phase separation \cite{CatesTailleurPRL,FilyPRL, rednerPRL,GomperWinkler,CatesTailleurRev,joakim_prl,lowen_CahnHilliard_prl} 
 in (quasi) two (in bulk or under confinment~\cite{andreasprl2013})
and three dimensions. }

\textcolor{black}{
However, when switching on an inter-particle attraction,   the phase behaviour of
 a dilute suspension of  attractive spheres  is still a matter of debate\cite{focus}.
Attractive self-propelled spheres  aggregate either into a 
network-forming phase, characterized by a  local alignment (without  aligning interactions)~\cite{filion} at high density, 
or into  a (steady state) cluster phase at low densities \cite{rednerPRE,mognetti}.
  The latter has been also observed in a dilute suspension of self-propelled spheres interacting via an isotropic 
short-range attraction and long-range repulsion\cite{dumbbells_microphase}.}

  \textcolor{black}{
When considering the effect of hydrodynamic interactions, numerical studies of a two dimension  suspension of self-propelled repulsive spheres  have
shown that hydrodynamics suppresses motility induced phase separation in dense suspensions\cite{ricardPRE2014},  
as suggested in the early work by Ishikawa \cite{IshikawaBandsPRL08}, and 
has an effect  in the dynamics of transient clusters at lower densities \cite{Llopis2DclustersEPL06}.}

\textcolor{black}{
Particle's shape has a strong impact in the collective behavior of suspensions of self-propelled particles. It has been shown that 
 repulsive active rods exhibit a phase-separated state characterized by the formation of polar clusters~\cite{weitzSPR}. 
 Similarly, dilute suspensions of  active dumbbells form rotating clusters when particles interact both via 
 a short-range attraction and long-range repulsion \cite{dumbbells_microphase} and 
 an isotropic attraction  \cite{pnas_chantal}, in the latter case displaying a nematic order with spiral patterns 
 in two dimensions \cite{suma_dumbbells}.}

\textcolor{black}{Inspired by experiments on dilute suspensions of  colloids \cite{Palacci13} and bacteria \cite{pnas_chantal},  
showing cluster formation, and  motivated by understanding how hydrodynamics could affect the 
  formation of living clusters \cite{dumbbells_microphase,mognetti,redner_softmat2013},} we have carried out a systematic numerical analysis on dilute 2D suspensions 
of attractive  self-propelled spheres both in the presence \textcolor{black}{(wet active system)} and absence \textcolor{black}{(dry active system)} of hydrodynamic interactions \cite{reviewMarchetti}. 
We identify the conditions at which the system forms clusters, and 
characterize their morphology considering and neglecting hydrodynamics. 
The comparison will provide further insight into the nature of the cluster phases in \textcolor{black}{  systems of attractive self-propelled spheres} and their main features.
Moreover, our study underlines the relevance of active stresses in the collective motion  of active suspensions \textcolor{black}{of spheres with attractive interactions}. 

 We perform  Lattice-Boltzmann simulations of  micro-swimmers 
 modelled  as squirmers~\cite{Lighthill}  
(to reproduce the   induced hydrodynamic flow around a spherical swimmer 
while preserving the main features of the active stresses generated by it~\cite{sunil}).
Both the  character of hydrodynamic active stress, either pushers or pullers,  $\beta$ (entering through   its sign) 
and the magnitude of the swimmer activity 
   affect the qualitative behaviour of a squirmer suspension, 
    as already  shown  for squirmers  close 
  to a solid wall~\cite{llopis_wall} and when studying their rheological properties~\cite{llopis_rheo}. 
Tuning both activity and attraction we are able to characterize different types of collective behaviour, 
distinguishing pushers from pullers (since depending on the hydrodynamic signature 
hydrodynamically interacting squirmers  can develop long-time polar order~\cite{alarcon2013}) and 
quantify to what extent  an explicit attraction between particles can affect their phase behaviour. 
To compare with the analogous system without hydrodynamics, we perform Brownian dynamics 
simulations of a 2D suspension of active Brownian disks (indicated as ABP in the text) \cite{mognetti,filion} .  
Differently from Ref. ~\cite{ricard_softmat2015},  on the one hand we  
take into account hydrodynamics  in three dimensions (or quasi 2D);  
on the other hand,  in the absence of hydrodynamics, we  consider both translational and rotational diffusion.

The manuscript is organized as follows. In section \ref{Sec:NumericalDetails} we present both the  
squirmer  
and the \textcolor{black}{spherical} active Brownian disk models. Next,  we describe the interaction potential and  introduce the 
parameters and analysis tools  used to characterize the different  collective behaviour. 
In section \ref{results} we present our results, analysing the dynamics
 of cluster formation 
  and discussing the different   clustering regimes depending 
 on the interaction strength versus activity. 
Finally, we discuss our conclusions in section \ref{Sec:conclusions}.

\section{Simulation details}\label{Sec:NumericalDetails}

\subsection{Squirmers}
We now   introduce a  hybrid scheme which resolves individual squirmers and the
   corresponding motion of the embedding fluid in a quasi two dimensional set-up. 
   This approach allows  us to  consistently follow the  dynamics of the fluid and the suspended squirmers on the same footing.

According to the squirmer model (Appendix~\ref{appx:squirmermodel}) 
 the solvent velocity, ${\bf u}|_{R_p}$, on the surface of a spherical  squirmer of radius $R_p$ can be expressed as
\begin{align}\label{surface_veloc}
 \textbf{u}|_{R_p} = \left[ B_1 \sin \theta + B_2 \sin \theta \cos \theta \right] \boldsymbol{\tau} ,
\end{align}
where $\boldsymbol{\tau}$ is a unit vector tangential to the particle surface and 
 the squirmer moves along the direction  ${\bf e}_1$  rigidly bound to the particle. 
  identifying an intrinsic axis from which the polar angle  $\theta$ can be determined.
 
 We restrict ourselves to the simplest squirmer model where the slip velocity depends
  only on two parameters:  $B_1$ that quantifies the asymptotic self-propelling speed, $v_s= \frac{2}{3} B_1$, at 
  which an isolated squirmer will swim, and $B_2$, proportional to the active stress generated by the squirmer.
  The ratio between the self-propelling velocity and active stress, $\beta=B_2/B_1$~\cite{Ishikawa}, 
  quantifies the active state of the squirmers and their interaction with the fluid. 
    $\beta >0$ corresponds to pullers, 
    generating a thrust in front of   their body, differently from pushers 
      ($\beta<0$) where the thrust is generated at their back  
  ~\cite{Ishikawa07}. 
  To simulate  squirmers we disregard thermal fluctuations: 
velocity fluctuations are simply induced by the particles' activity  ($B_2$) 
 acting as an effective temperature when competing with conservative forces. 

   The  surrounding fluid is modelled using  a Lattice Boltzmann (LB) approach. 
The solvent is described in terms of the one-particle  distribution function, $f_i({\bf r},t)$ 
i.e. the density of a particle with velocity ${\bf c}_i$ at a node (${\bf r}$) of a given lattice. 
The discretized velocities join nodes  and prescribe the lattice  connectivity. We  use the D3Q19 lattice, 
characterized by 19 velocities joining nodes of a cubic three dimensional lattice~\cite{cates_lb}. The  
fluid dynamics emerge from the  evolution of the one-particle distribution  function,
\begin{equation}
f_i({\bf r}+{\bf c}_i, t+1) = f_i({\bf r}+{\bf c}_i, t)+\Delta_i({\bf r},t)
\end{equation}
where $\Delta_i$ can be understood as the linearised Boltzmann collision operator that relaxes the 
densities toward a local equilibrium and conserves mass and momentum. This LB model reproduces the 
dynamics of a Newtonian liquid of shear viscosity $\eta$ and the relevant hydrodynamic variables are recovered as moments of the 
one-particle velocity distribution functions~\cite{succi}.

In our simulations particles  (with a diameter $\sigma$ of 4.6 lattice nodes~\cite{nguyen})   are individually resolved,  
 imposing a modified bounce back rule for spherical colloids on the  one-particle velocity distribution functions  for  nodes  
 crossing the solid interface  moving from a node outside to one inside the particle
 (including the slip velocity
 to impose  the squirming motion~\cite{llopis_wall}).
   The total force and torque the fluid exerts on the particle is obtained by
  imposing that the total momentum exchange between the solid particle and the fluid nodes vanishes
    (as a result of   the modified bounce back) . Accounting for all forces acting on each squirmer allows to update 
  them dynamically. In particular, the torque exerted by the fluid determines how the squirmer direction of  
  selfpropulsion, ${\bf e}_1$ will rotate.
   In our work we have used  a LB code which consists of moving 
    particles via a domain decomposition  and parallelization (using MPI) \cite{ludwig} 
    to exploit the excellent scalability of LB on supercomputing facilities. 

We simulate attractive squirmers  via a truncated and shifted Lennard-Jones potential, of magnitude
\begin{equation}\label{eq:apolaratt}
\begin{tabular}{ r c l }
 \(  V(r) \) & \(=\) & \( 4\epsilon \left[ \left(\frac{\sigma}{r}\right)^{12} - \left(\frac{\sigma}{r} \right)^6 \right] \);
\end{tabular}
\end{equation}
when two squirmers are at a  distance $r$ smaller than $r_{cut}= 2.5 \sigma$, and vanishing otherwise. 
The competition between attractive and self-propelling forces  is quantified through the dimensionless parameter
\begin{equation}
\xi = \frac{F_d}{F_{LJ}},
\end{equation}
where $F_d = 6 \pi \eta R_p v_s$ is a characteristic magnitude of the  friction force associated to the squirmer self propulsion and 
$F_{LJ}$ is the absolute value of the of Lennard-Jones force at its minimum $r=(26/7)^{1/6}\sigma$.

We will analyse the dynamics of squirmer suspensions in quasi 2D system  ($L\times L \times k \sigma$)  
 with periodic boundary conditions. 
 While LB captures the three dimensional hydrodynamics associated to the system geometry, 
 squirmers  are confined to move on a plane, ensuring that both the component of the velocity perpendicular to the plane 
and the angular velocity  acting off-plane, vanish.
For computational convenience, 
the thickness of the slab is larger than a particle diameter ($k=5$).   
We will consider the collective dynamics of   semidilute suspensions  at $\phi=0.10$ (where $\phi=\frac{\pi}{4} \rho\sigma^2$ and  $\rho=N/L^2$ ) 
and simulate  $N= 10 000$   \textcolor{black}{spheres} 
to minimize finite size effects   (corresponding to lateral box sizes of $L \approx 111 r_{cut}$).
(As a double check, we have also run simulations for larger systems and have not observed
significant deviations).
A squirmer travels its own size in $t_0 = \frac{\sigma}{v_s}$. We have run simulations from $1450$ up to $3000$ $t_0$. 
 Once the system reaches steady state, in a time window between  
 $1000$ and  $2000$ $t_0$, we 
carry out a systematic analysis of the dynamics of the squirmer suspension,  considering  $\xi$  from 1 to $\infty$ and $\beta$ from -3 up to 3. 

\subsection{Active Brownian Disks}
\label{brownianSection}
For the two dimensional  Brownian dynamics simulations, each of the $N$ active disks (with diameter $\sigma$) 
are represented by their positions 
and self-propulsion directions $\left\lbrace \textbf{r}_i, \theta_i\right\rbrace$,   
both satisfying the coupled overdamped Langevin equations,
\begin{equation}
\label{bro1}
\dot{\textbf{r}}_i=\frac{D}{k_BT}\left[\mathbf{F}_{LJ}\left(\left\lbrace\textbf{r}_i \right\rbrace \right) + F_p \bf{e}_i\right]+\sqrt{2D}\mathbf{\eta}_i^T,
\end{equation}
\begin{equation}
\label{bro2}
\dot{\theta}_i=\sqrt{2D_r}\eta_i^R,
\end{equation}
where $\mathbf{F}_{LJ}$ is the force due to the Lennard-Jones potential (eq.~\ref{eq:apolaratt}), 
$F_p$  the magnitude of the self-propulsion  which, in the absence of interactions, will move a particle with speed  $v_p=\frac{D}{k_BT} F_p$  
 and  $\bf{e}_i$ = $(\cos \theta_i, \sin \theta_i)$. $D$ and $D_r$ are translational and rotational 
diffusion constants, which in the low-Reynolds-number regime can be related 
by $D_r=3D/\sigma^2$. The $\eta_i^{T,R}$ are Gaussian white noise variables with 
$\langle \eta_i(t) \rangle = 0$ and $\langle \eta_i(t)\eta_j(t') \rangle=\delta_{ij}\delta(t-t')$.
 \textcolor{black}{Equations \ref{bro1} and \ref{bro2} are tailored for Brownian disks, where
  the direction of the self-propulsion  ($\theta_i$) does not induce a torque (in contrast with the dynamics for Active Brownian Rods~\cite{weitzSPR}).} 
  
The Brownian dynamics  for \textcolor{black}{disk shaped} Active Brownian particles (ABP) we implement evolves in time according to
an Euler scheme with a time step of $5$ $10^{-5}$ $\tau$, being $\tau$ the time unit 
($\sigma^2/D$, where $D$ is set to 1 $\sigma^2/\tau$), $\sigma$ the length unit and  $k_B T$ the energy unit. 
The reduced temperature ($k_BT/\epsilon$) is set to 0.2 (being $1/k_BT=1$). 
Brownian dynamics simulations are run over $10^8 \tau$, and the time a particle travels
 its own size is $t_0=\sigma/v_p$. We have started measuring only when the system reached a clear steady state.

We  have considered the semidilute regime, and analyse the behaviour of a suspension at $\phi=0.10 $. 
To minimize finite size effects, we have simulated N = 10000 self-propelled \textcolor{black}{disks}.  
When setting the propulsion force to zero, we recover the results expected for a dilute suspension 
of Lennard-Jones passive particles interacting via eq. 
\ref{eq:apolaratt} as in Ref.~\cite{smit} 
with the same parameters ($\phi=0.$1 and the reduced temperature $0.2$).
At the chosen temperature and density, the equilibrium counterpart of our system  lies in the liquid-vapor coexistence region and 
forms a steady state distribution of small clusters.
When simulating the passive Brownian case ($F_p=0$), we also recover the results obtained by 
Matas-Navarro and Fielding\cite{ricard_softmat2015} for a similar value of $\phi$ ($\phi=0.125$) and $\widetilde{D} = 0.2$: 
once more, the system forms a steady state cluster distribution.  

To quantify the competition between attraction and propulsion, we study the system at different values of $\xi = \frac{F_p \sigma}{\epsilon}$
(that corresponds to $P_{agg}^{-1}$ used in Ref.\cite{mognetti,filion})
 ranging from 0.5 up to  6, in order to be able to compare  with results obtained for squirmers .

\subsection{Analysis tools}

We identify clusters based on a distance criterion: two particles belong to the same cluster whenever their distance is smaller than 
$r_{cl}=1.8\sigma$ for squirmers and  $r_{cl}=1.5\sigma$ for ABP (see Appendix). 

We  evaluate the fraction of clusters of size $s$, $f(s)$ as a measure of the
 cluster-size distribution. To calculate $f(s)$ we apply the same criterion as in Ref. \cite{chantalbins2012}:
  (i) We arbitrarily subdivide the range of $s$-values into intervals $\Delta s_i=(s_{i,max}-s_{i,min})$, 
  where we define the total number of clusters within each interval $\Delta s$ as $n^t_i$; (ii) 
   we assign the value $n_i=n^t_i/\Delta s_i$ to every $s$ within $\Delta s_i$, and compute the fraction 
   of clusters of size $s$  as $f(s)=n_i/N_c$ where $N_c=\sum_i n_i \Delta s_i$ 
is the total number of clusters. 

 We  also compute morphological and orientational properties of the clusters, such as  their  radius of gyration\begin{equation}\label{eq:R_g}
R_g = \sqrt{\sum_{i,j=1}^{s} \frac{(\mathbf{r_i} - \mathbf{r_j})^2 }{2 s}}
\end{equation}
(dividing by 2 to avoid double counting) and 
their  polar order 
\begin{equation}\label{eq:P}
P = \left|  \frac{\sum_{i=1}^{s}  \mathbf{e}_i }{s} \right|.
\end{equation}
The cluster's orientation with respect to the cluster's translation ($\mathbf {v}_{CM}$) is also determined  
\begin{equation}\label{eq:Omega}
\Omega (s) =  \frac{\mathbf {v}_{CM}(s) \cdot \mathbf{P}(s) }{{v}_{CM}P}
\end{equation}
where $\mathbf {v}_{CM}(s)$ is the cluster center of mass velocity, and $\mathbf{P}(s)$ is defined as
\begin{equation}
\mathbf{P}(s) = \frac{\sum_{i=1}^{s}  \mathbf{e}_i }{s}.
\end{equation}

\section{Results}\label{results}
\subsection{Mean cluster size}
\begin{figure}[h!]
\begin{center}
\includegraphics[clip,scale=0.6]{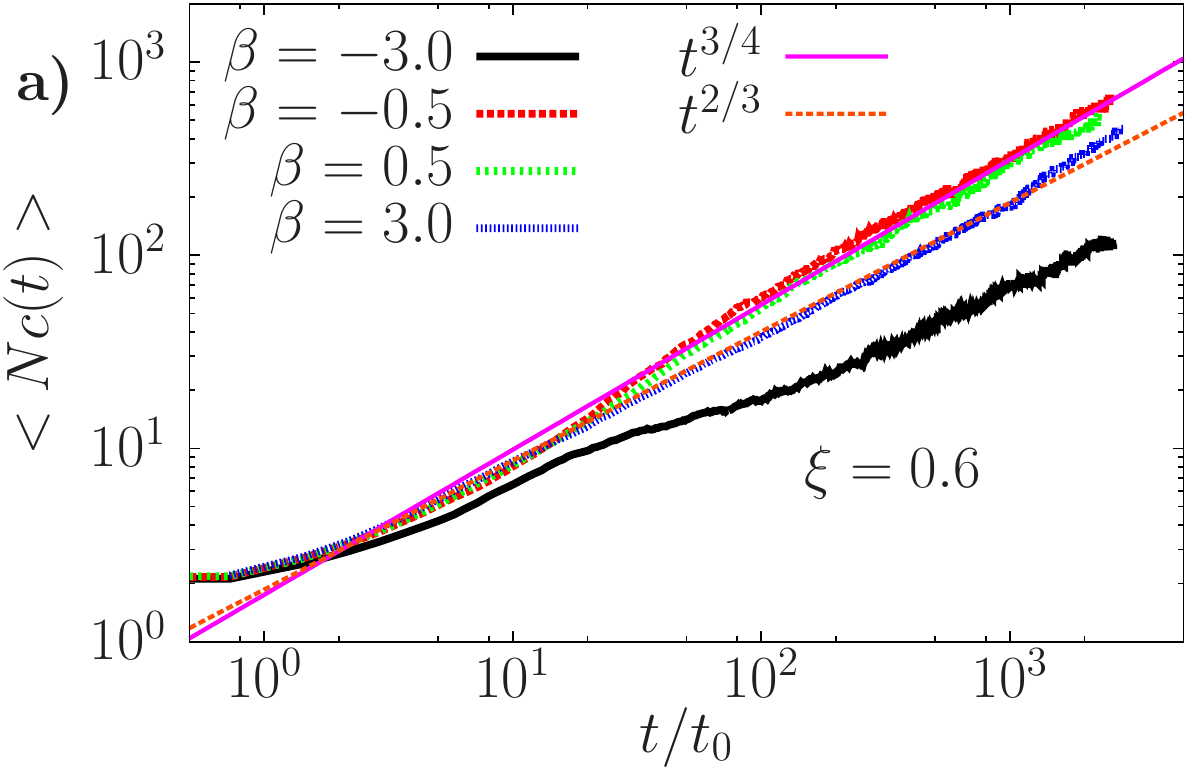}\\
\includegraphics[clip,scale=0.6]{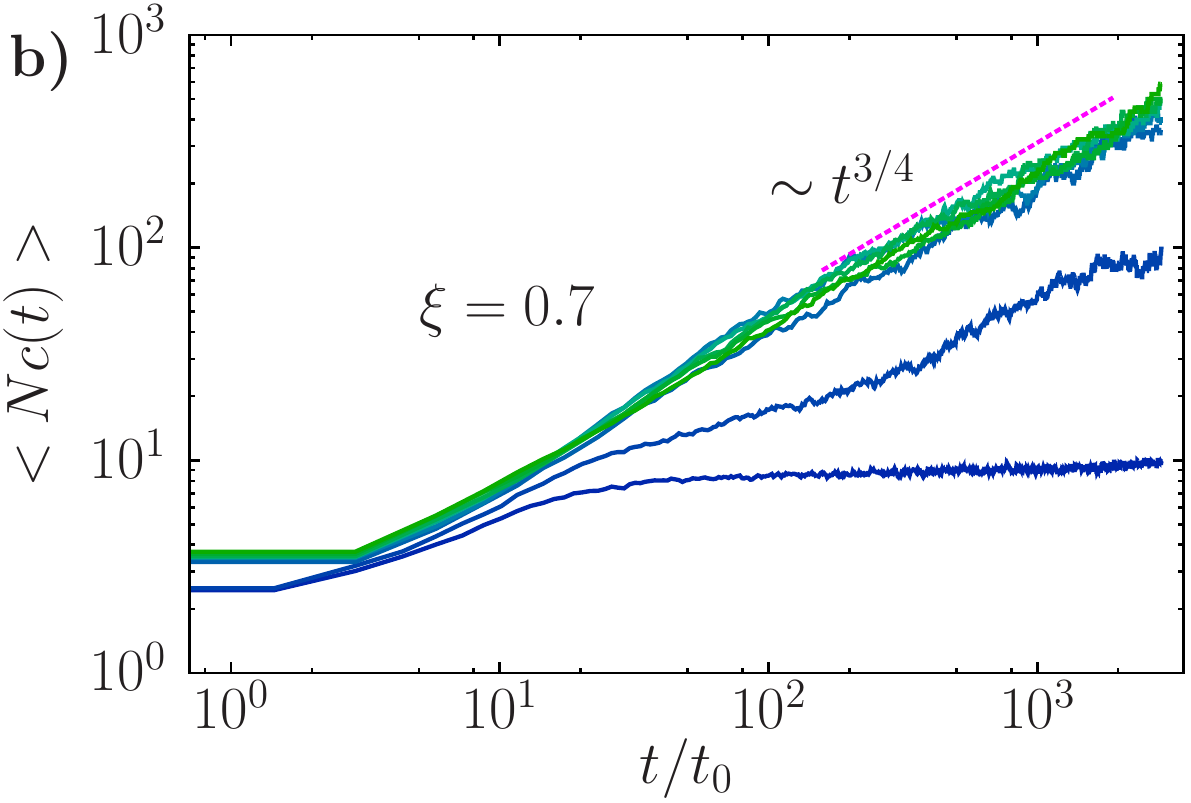}  
\caption{a) Mean cluster size as a function of time for squirmers with strong LJ interactions:  a) $\xi=0.6$ and  b) $\xi=0.7$. 
In both panels dashed orange and solid pink curves represent $\sim t^{2/3}$ and $t^{3/4}$ respectively and active stress  $\beta$ ranges from $-3$ to $3$. 
}
\label{Ncmean_xi0_6}
\end{center}
\end{figure}

In order to assess when the suspension reaches a  steady state~\cite{alarcon2013}, 
we have computed the time dependence of the mean cluster size for the squirmer suspension, 
as both  the mean cluster size and the global polar order must saturate at  steady state. 

The analysis of the time evolution of these quantities has allowed us to identify three different  scenarios. 
As shown in Fig.~\ref{Ncmean_xi0_6}-a for $\xi = 0.6$  (strong LJ interactions,   $\xi < 1$) squirmer suspensions display 
coarsening. In this regime attraction dominates over self-propulsion leading  to cluster growth. Pushers coarsen faster than pullers due to an intermediate slowdown for the latter. In both cases, the asymptotic coarsening is compatible with an algebraic growth, with exponent $3/4$, larger than the previously reported asymptotically  coarsening  exponent, $\alpha=2/3$, on 2D attractive squirmers at $\beta=0$~\cite{ricard_softmat2015}.

Increasing $\xi$ even further, we find a regime where coarsening depends 
on the active stress. This regime is shown in  Fig. \ref{Ncmean_xi0_6}-b for $\xi=0.7$.  
This is a clear  example where the collective  behaviour of the active suspension 
 depends not only on its degree of activity, but also on the type of hydrodynamic stresses 
 squirmers induce in the surrounding fluid.  As we decrease $\xi$, the range of $\beta$ 
 for which coarsening is observed increases asymmetrically between pushers and pullers.  

\begin{figure*}[h!]
\begin{center}
\includegraphics[clip,scale=0.5]{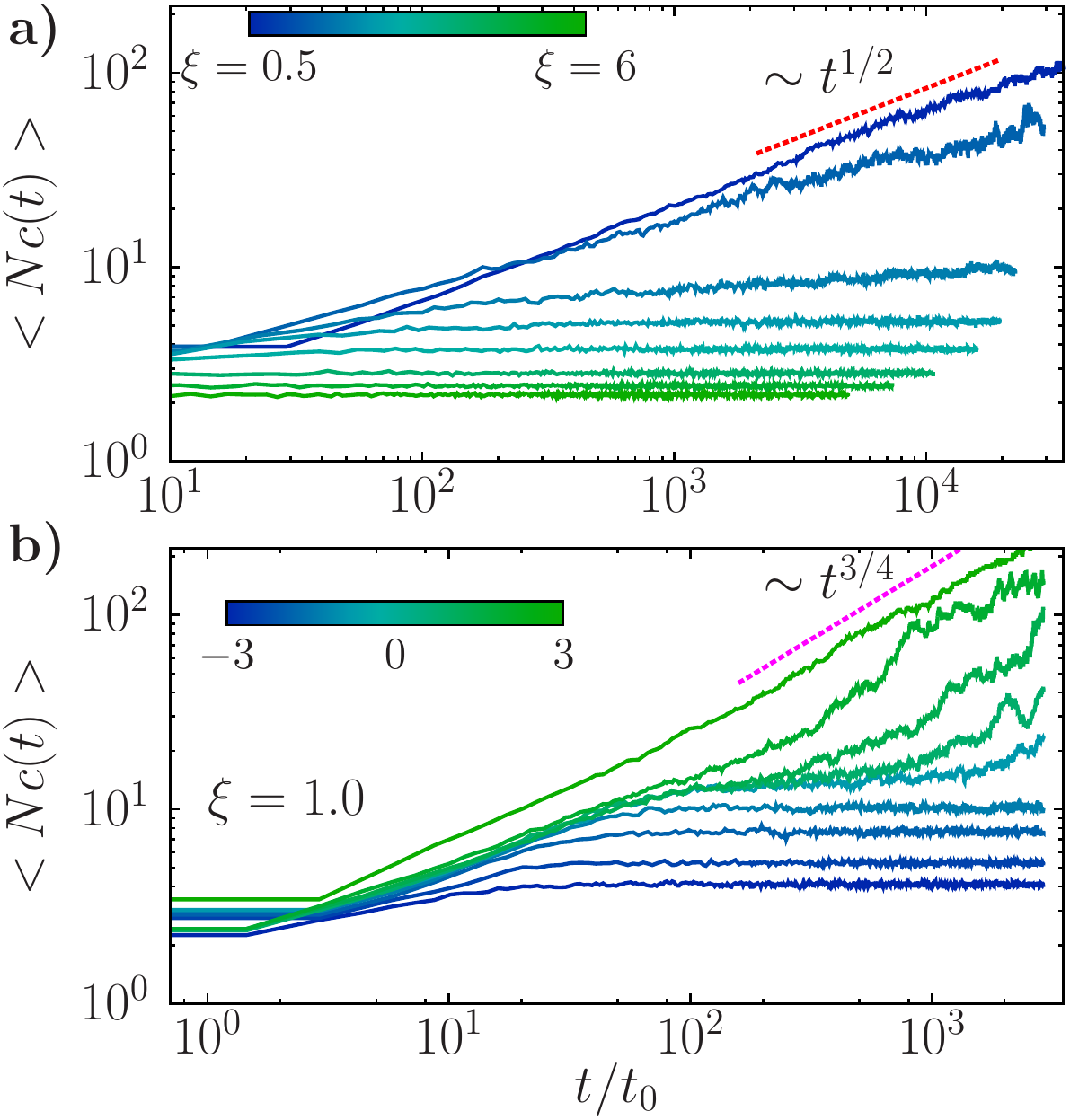} 
\includegraphics[clip,scale=0.62]{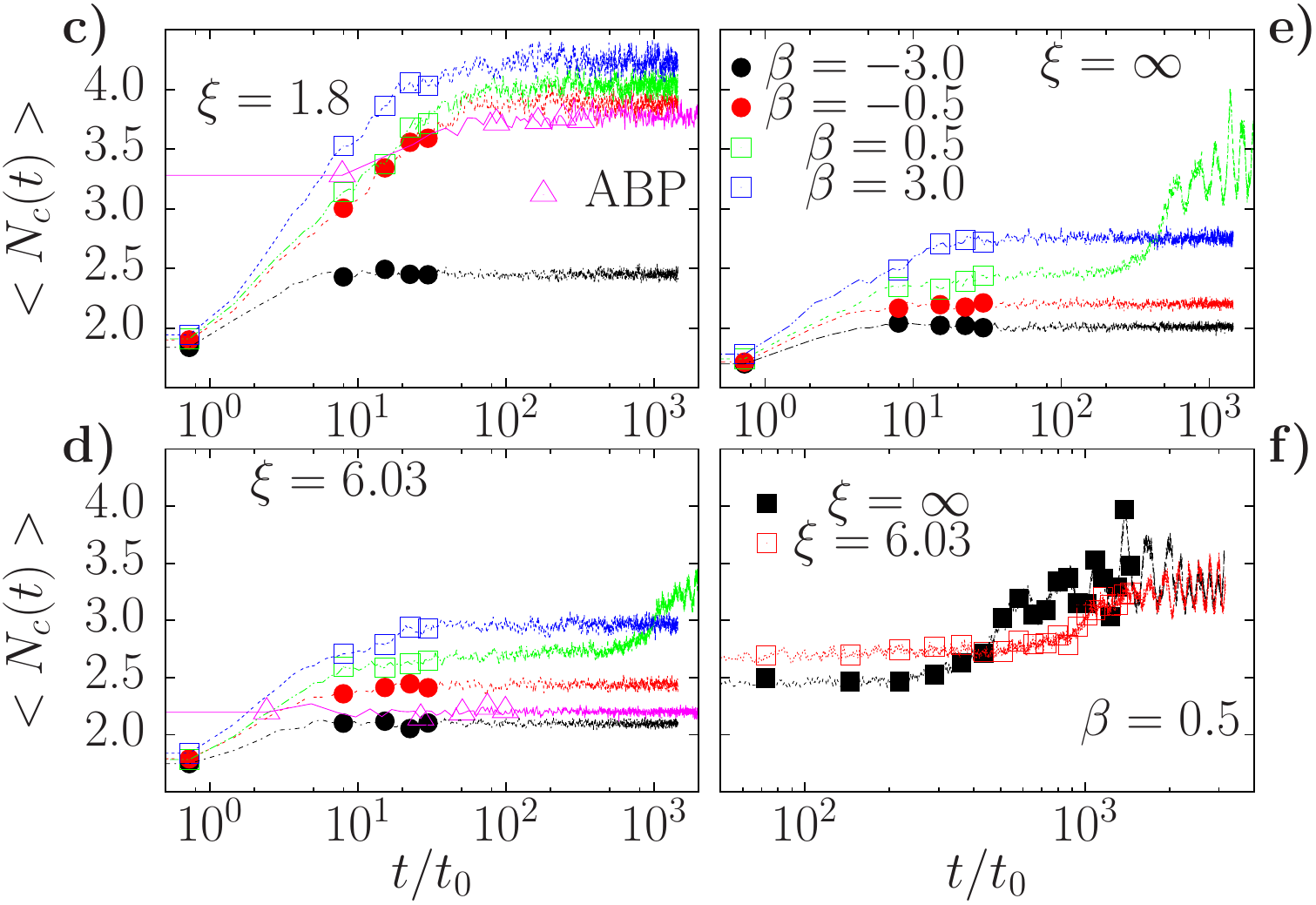} 
\caption{Mean cluster size as a function of time for different cases. (a) Brownian \textcolor{black}{spherical} particles at different values of interaction strength, from $\xi=0.5$ to $6$. The red line corresponds to the curve $\sim t^{1/2}$. (b) Squirmers with $\xi=1$ for various values of $\beta$ from $-3$ to $3$. The pink line correspond to the curve $\sim t^{3/4}$. (c) Squirmers with $\xi=1.8$. Circles correspond to $\beta <0$ whereas squares to  $\beta >0$, pink triangle correspond to ABP with the same interaction strength. (d) Squirmers and ABP with  $\xi=6.03$. (e) Squirmers with $\xi=\infty$ and (f) mean cluster size for $\beta = 0.5$ with $\xi=6.03$ and $\xi=\infty$ up to $t/t_0=3000$.}
\label{Ncmean}
\end{center}
\end{figure*}

Figure \ref{Ncmean} summarizes all  results obtained  for ABP and  squirmers. 
In both cases, cluster formation is due to the competition between attraction and self-propulsion, as quantified by $\xi$.

Figure~\ref{Ncmean}-a displays  the mean cluster-size as a function of time for self-propelled \textcolor{black}{spherical} Brownian particles in two dimensions at different interaction strength.  
  When attraction is stronger than propulsion ($\xi=0.5, 1.0$ and $1.3$, dark blue curves),
  we observe the system coarsening and that 
   the stronger the attraction the faster the coarsening.  
 Whereas when attraction competes with propulsion ( $\xi>1.3$) the system quickly enters 
 a  regime of steady state clusters, whose size depends on the propulsion strength 
 (the higher the propulsion, the smaller the clusters).

   In Figure \ref{Ncmean}-b, we observe that 
the mean cluster size for $\xi=1$ is sensitive to the nature of the active  hydrodynamics 
interactions, and changes with  $\beta$: 
pushers  ($\beta < 0$) reach a steady state  faster and the larger the $|\beta|$ 
the lower the 
mean cluster size; whereas pullers, $\beta \geq 0$, coarsen. 
Pullers with $\beta < 3$ first reach a metastable mean cluster size of about 15 particles  
at short times, and their  mean cluster size grows again due to cluster-cluster interactions:
 the smaller the $\beta$  the longer the time  a suspension remains in its metastable state.

To ensure the robustness of either the cluster of the coarsening state,  
we have carried out simulations over longer times and  simulations whose 
initial configuration consisted of all particles forming one big cluster: we observed that while 
at the beginning the mean cluster size dropped to the metastable state,  then grew in the same way as when  
 particles were randomly distributed at the beginning of the simulation.

 Cluster formation is caused by attraction competing with self-propulsion. 
However, for squirmers active stresses modulate this competition. As a result, we observe that pullers promote cluster 
formation because of their tendency to align to nearby squirmers. 
Therefore, for a given $\xi$, a larger fraction of pullers  
coarsen compared to pushers of analogous active stress. 
For an equivalent $\xi$, clusters  of active Brownian \textcolor{black}{disks} are on average larger than squirmers one, indicating 
that   active stresses contribute to break clusters due to an induced  hydrodynamic dispersion. However, we 
cannot exclude that  by fine tuning  $\beta$ there may be ranges of $\xi$ for which the same mean cluster size can be 
obtained both for ABP and  squirmers.

As shown in Figure~\ref{Ncmean}-c-f, when  we further increase  the activity  beyond $\xi \geq 1.5$ 
we enter a  regime where all  suspensions reach a steady state in which squirmers do not 
merge in one cluster, and the mean cluster size decreases as  $\xi$  increases. 
When $\xi=1.8$ (Figure \ref{Ncmean}-c) coarsening disappears even for pullers, and a steady state cluster size 
distribution is reached for all squirmers. 
 ABP with the same $\xi$ develop a mean cluster size similar to the 
pushers with a slightly negative value of $\beta$ ($\beta = -1/2$). 
Therefore, we conclude that the stress activity dictates the magnitude of the average cluster size, since 
pullers form larger clusters than pushers. 

If we now increase $\xi$ even further (Figure \ref{Ncmean}-d), once
 more we observe all systems reaching a steady state cluster size distribution independently on the value of $\beta$. 
 The average cluster size is now smaller than for $\xi=1.8$. 
 For weak pullers ($\beta = 0.5$) the system exhibits a transient arrest:  
  this behaviour could be attributed to the development of metastable clusters,  
 before reaching the stable distribution for  longer times. 
 Moreover,  these puller suspensions exhibit stronger 
 density fluctuations (as we will discuss later) which could underlining  this peculiar behaviour. 
 Once more even for higher values of $\xi$,   ABP with $\xi=6.03$ 
  evolve to a mean cluster size similar to weak pushers. 
\begin{figure*}[ht!]
\begin{center}
\includegraphics[clip,scale=0.13]{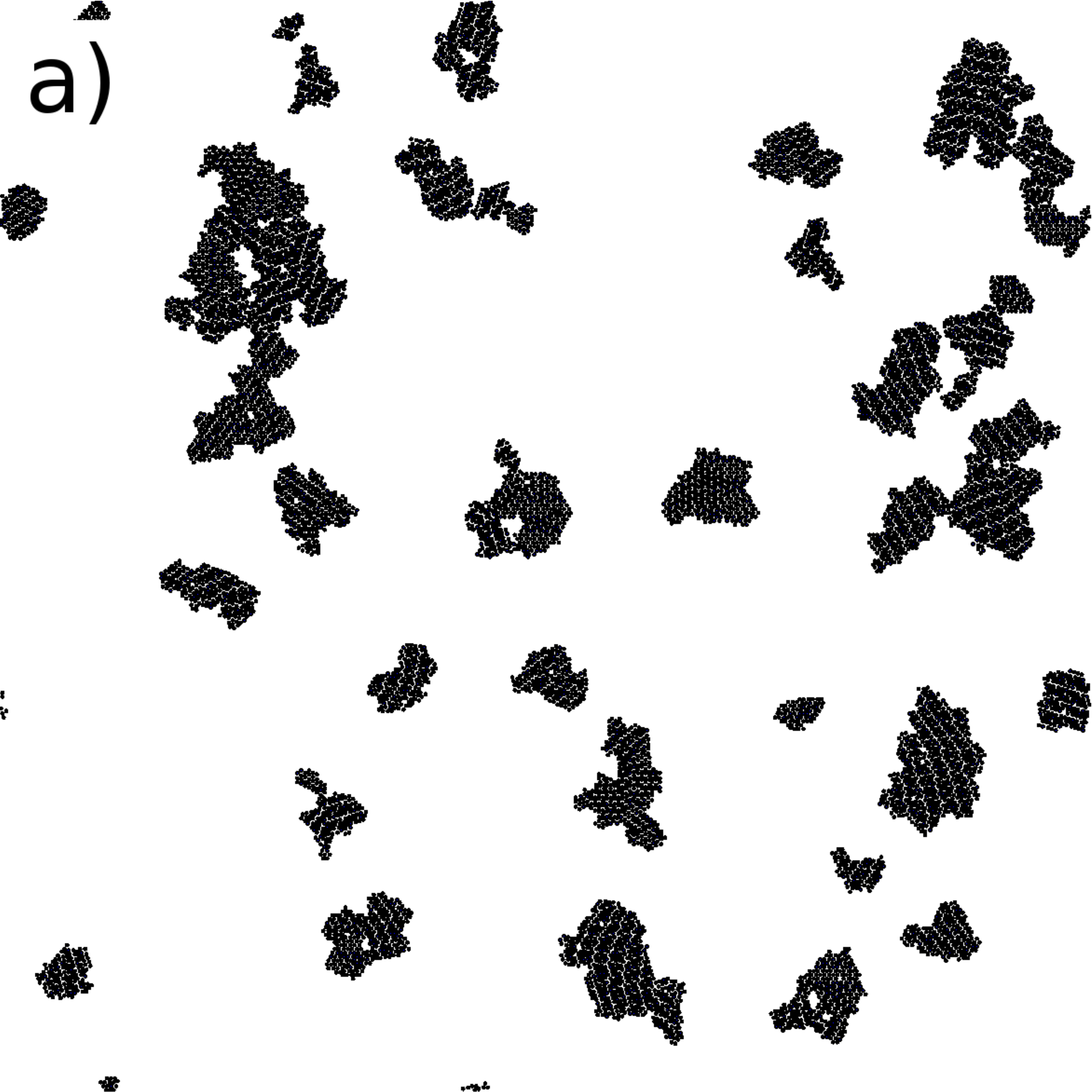} 
\includegraphics[clip,scale=0.13]{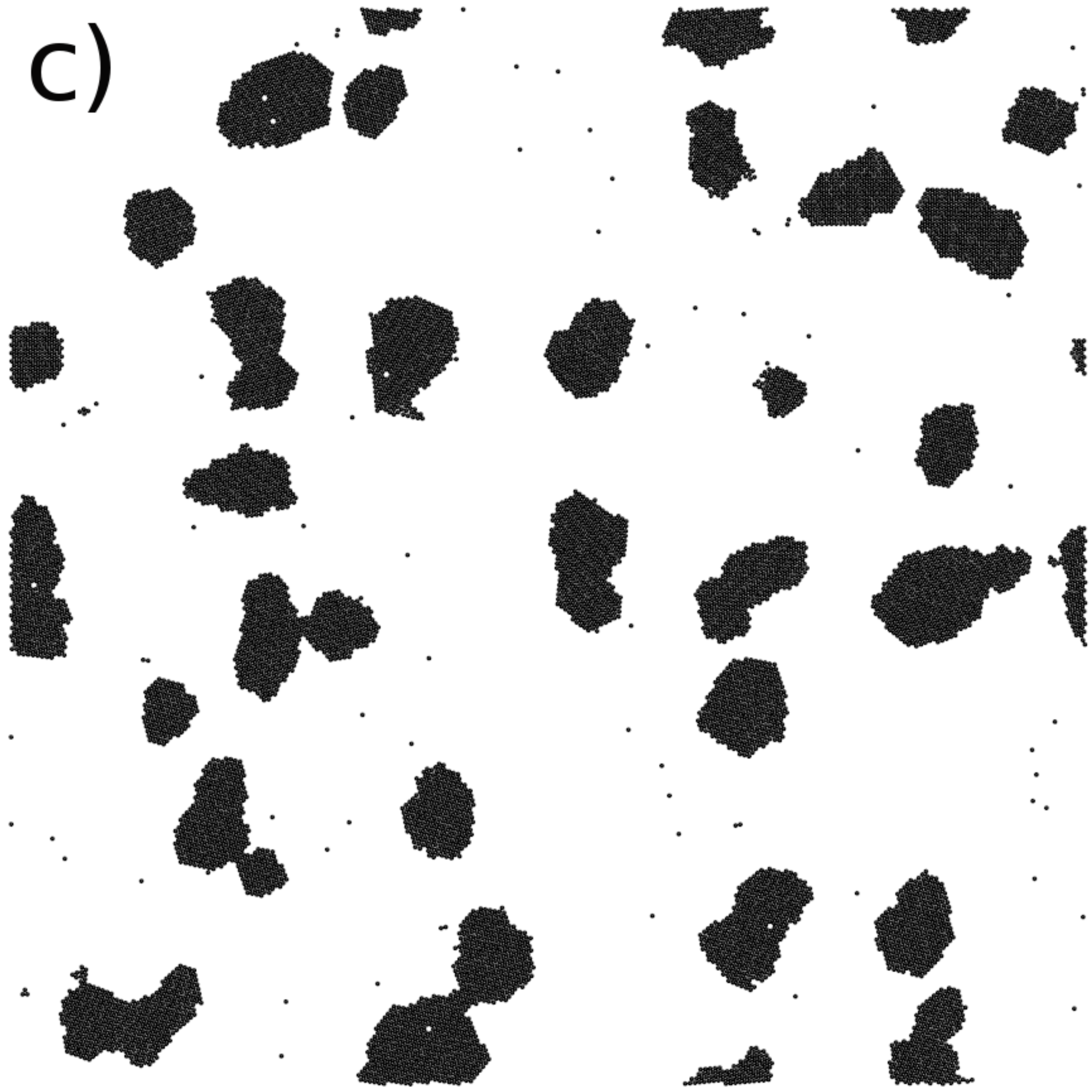} 
\includegraphics[clip,scale=0.13]{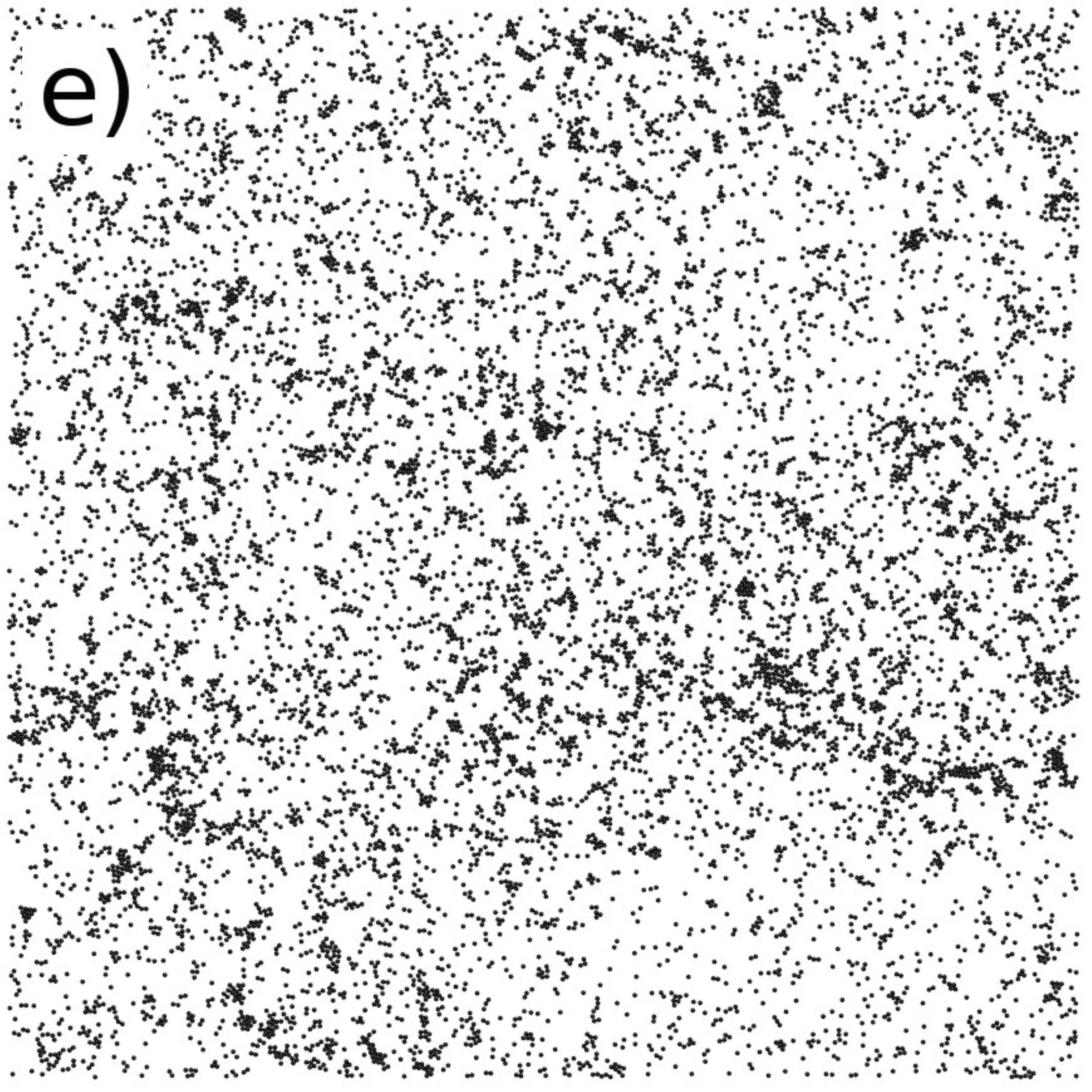} 
\includegraphics[clip,scale=0.13]{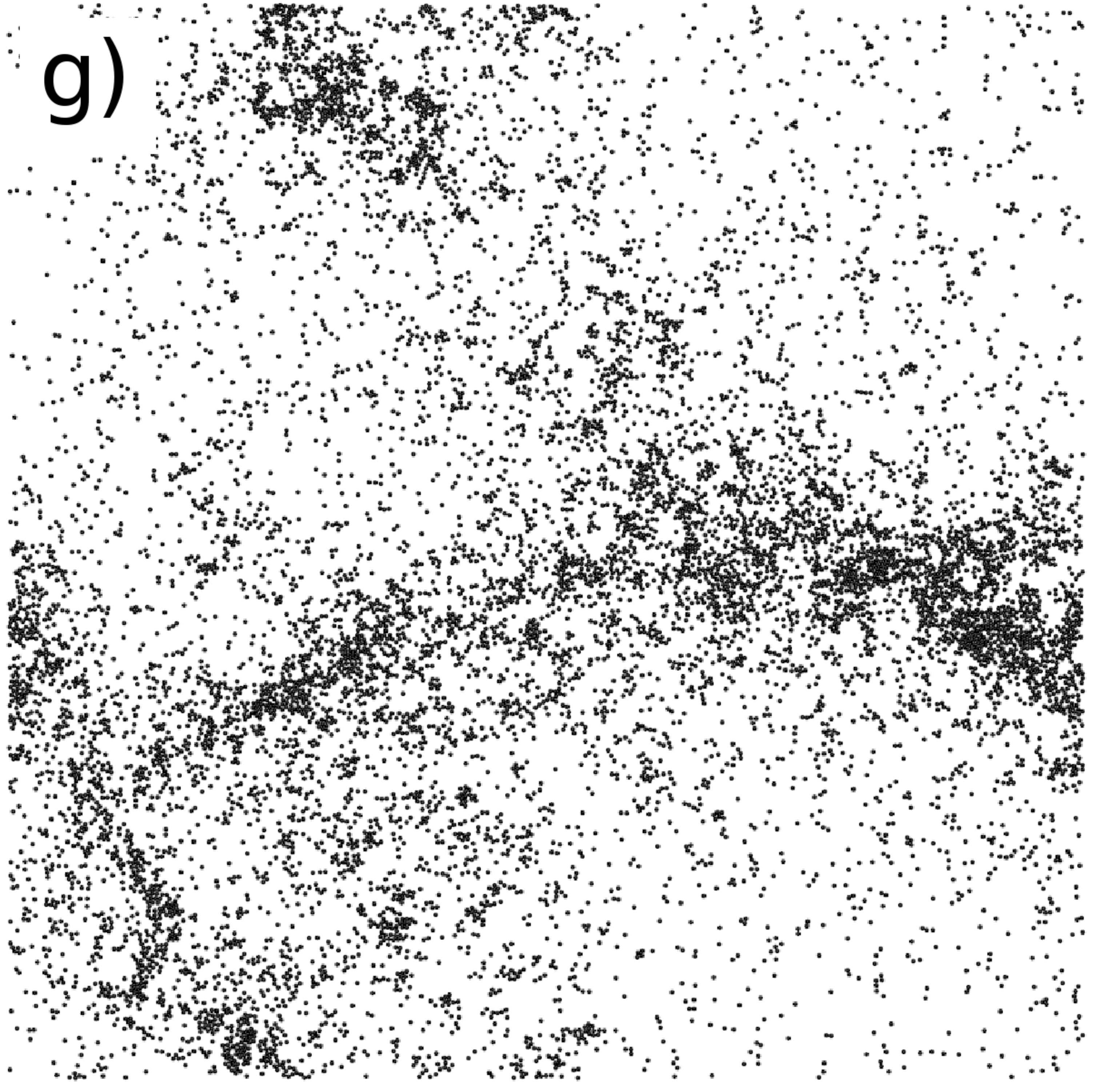} \\
\includegraphics[clip,scale=0.13]{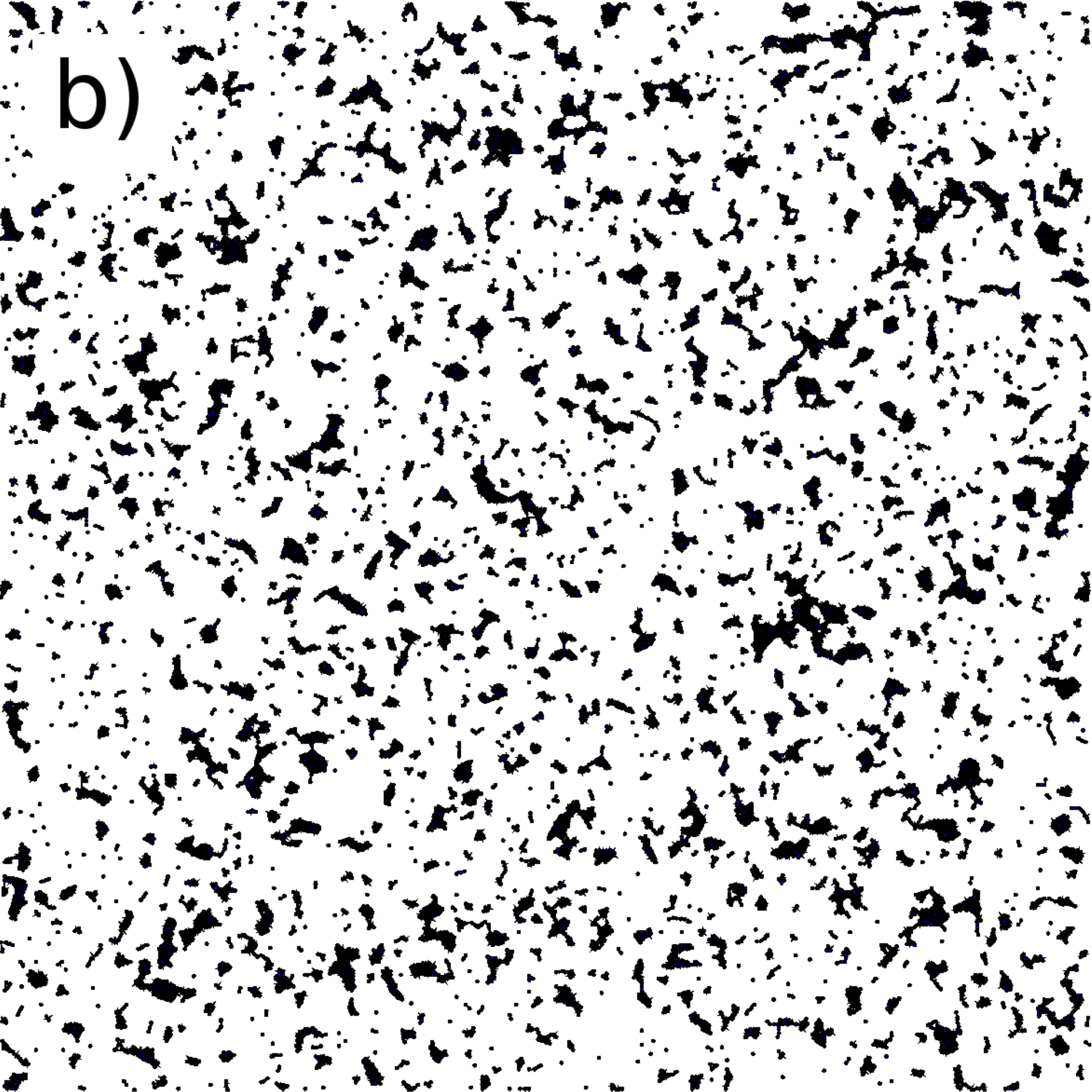}
\includegraphics[clip,scale=0.13]{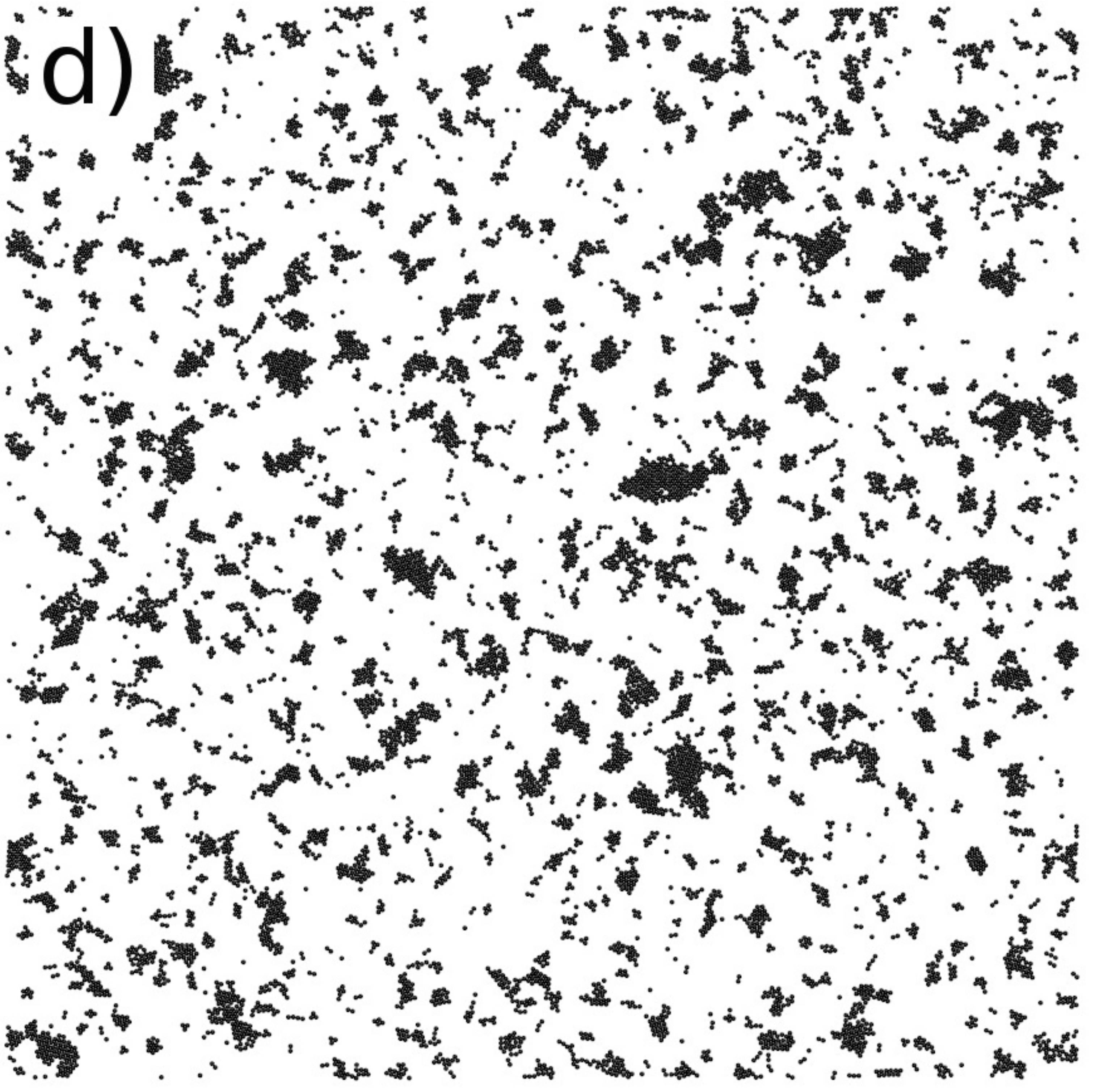}
\includegraphics[clip,scale=0.13]{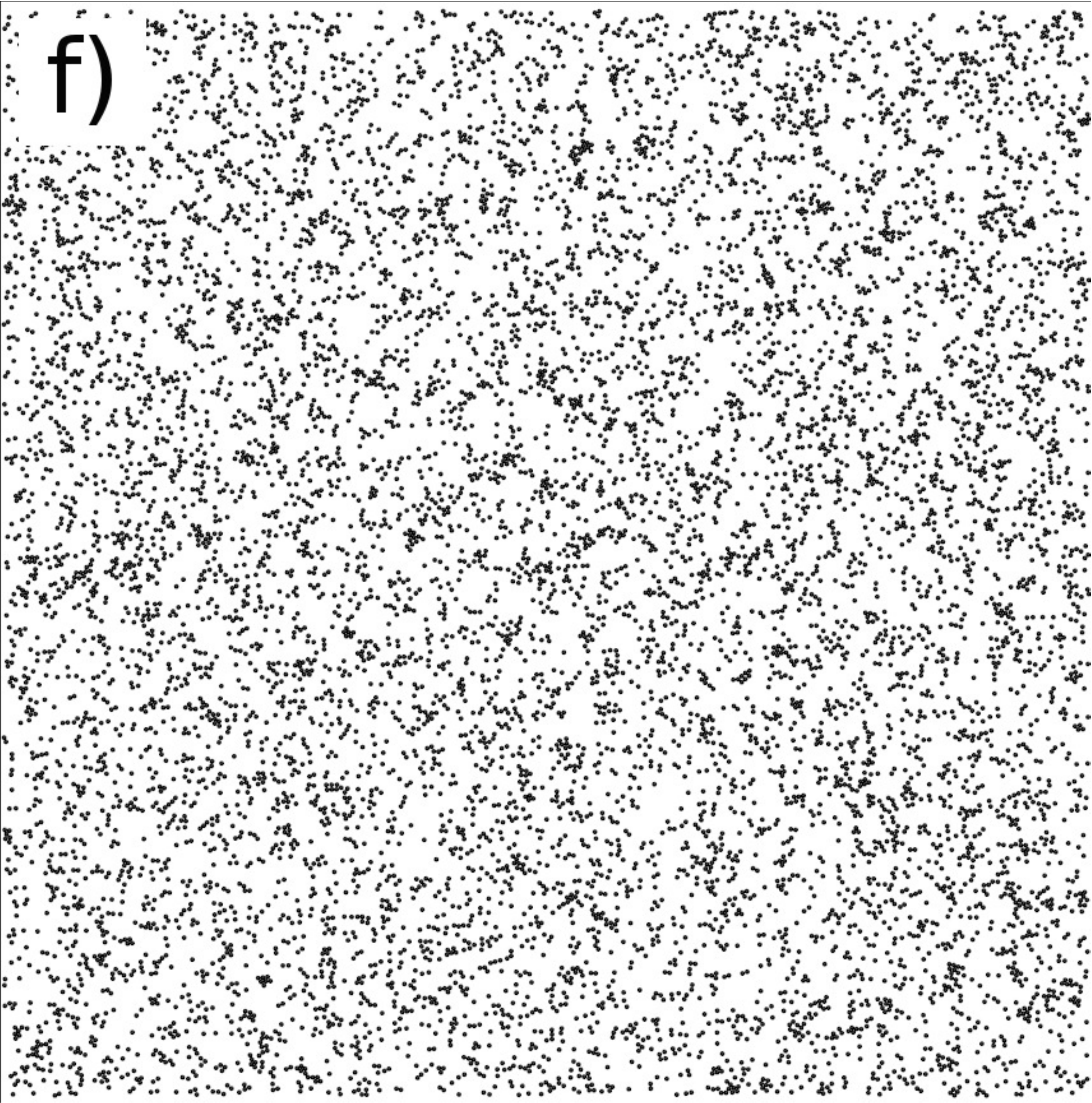}
\includegraphics[clip,scale=0.13]{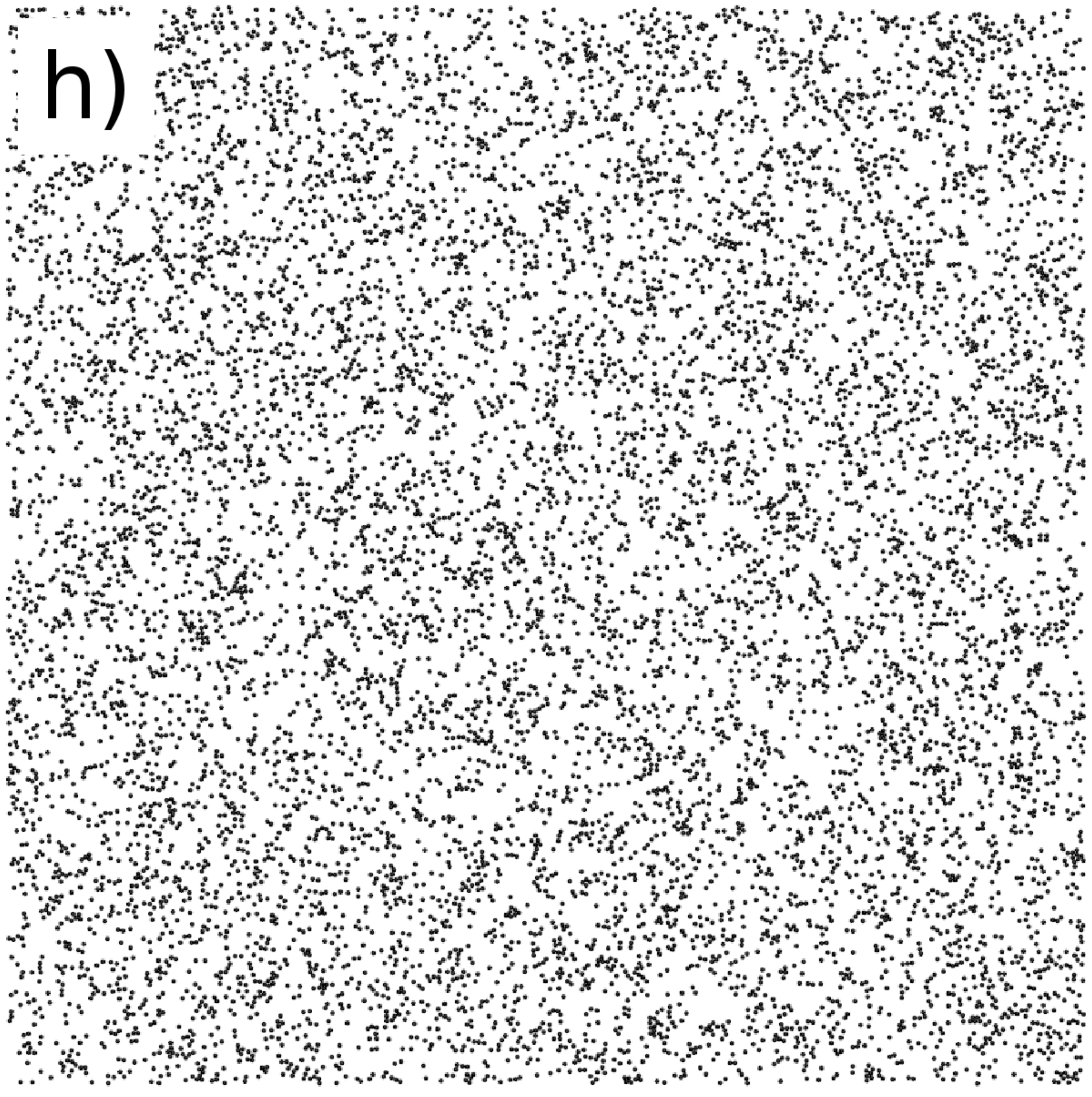}
\caption{Snapshots for different values of interaction strength for ABP and squirmers suspensions. 
The first column represent Active Brownian \textcolor{black}{spheres}, whereas all other columns are squirmers. a) 
ABP with $\xi=1$, b) ABP with $\xi=4$, c) Squirmers with $\xi=1$ and $\beta=0.5$, d) 
Squirmers with $\xi=1$ and $\beta=-0.5$, e) Squirmers with $\xi=5$ and $\beta=0.5$, f) 
Squirmers with $\xi=5$ and $\beta=-0.5$, g) Squirmers with $\xi=\infty$ and $\beta=0.5$ and h) 
Squirmers with $\xi=\infty$ and $\beta=-0.5$. All snapshots for squirmers were taken at $t/t_0 = 1450$. 
}
\label{photos}
\end{center}
\end{figure*}

 To conclude, we also analyse repulsive squirmers, $\xi=\infty$ (Figure \ref{Ncmean}-e and f) 
 where $<N_c(t)>$ behaves in the same way as for 
  $\xi=6.03$.
 We observe that when $\beta = 0.5$ and $\xi$  beyond  $\xi=6.03$  $<N_c(t)>$ oscillates at long times  (beyond the meta-stable state), 
 and identify this feature as a third scenario where the mean cluster size of the suspension has an oscillatory 
 behaviour, associated  with a specific parameters range (for $\xi \gg 1$ and $0<\beta<1$).
   To validate the persistence of the fluctuations,   we  run these simulations for  longer time 
   and conclude that     
 1) the  attraction strength  only affects the relaxation time towards this oscillatory regime, 
  2) the larger the value of $\xi$ the smaller the relaxation time.  
  Therefore,   while the stress generated by the squirmers promotes both 
  alignment and clustering between them, favouring   oscillations of the mean cluster size, the attractive 
  interaction partially competes delaying the relaxation to the regime controlled by active stresses.

In Figure \ref{photos}  we represent instantaneous snapshots for \textcolor{black}{spherical} attractive ABP with $\xi=1$ (Fig. \ref{photos}-a) and $\xi=4$ (Fig. \ref{photos}-b) as
 well as several regimes of  squirmers with $|\beta = 0.5|$  (Figures \ref{photos} c to h).
For $\xi=1$ ABP coarsen, whereas for $\xi=4$ a steady state clustering is observed. For squirmers  richer behaviour is 
observed depending on the hydrodynamic character of their active stresses, since for $\xi=1$ we observe either 
coarsening if $\beta=0.5$ (Fig. \ref{photos}-c) or clustering if $\beta=-0.5$ (Fig. \ref{photos}-d).
Even though coarsening disappears increasing  $\xi$  both for squirmers and for ABP, particles are not distributed in the same way 
(as shown in Figures \ref{photos}-b and \ref{photos}-e and f) and squirmer  clusters are smaller, on average, 
than their Brownian counterparts. 
  Comparing figure \ref{photos}-g and \ref{photos}-h, one  observe how pullers form clusters of particles moving 
 in the same direction  (figure \ref{photos}-g) , that percolate when the collective behaviour only depends on the stress activity.   
 On the contrary, for pushers the suspension is completely homogeneous (Fig. \ref{photos}-h).
 \textcolor{black}{ In the Supplementary information we have included representative movies to better capture the features reported in Fig. \ref{photos}.}

\subsection{Density fluctuations in squirmer suspensions}

The nature of the emerging clusters and their  intrinsic dynamics  can be indirectly measured through the 
number density fluctuations, evaluating  the mean number of particles $\overline{N}$ and 
the standard deviation $\Delta N$ 
as a function of the size of the  subsystem in 
which we partition the suspension.
Asymptotically, the standard deviation obeys  to a power law, $ \left\langle \Delta N \right\rangle \sim N ^\alpha $. 
If fluctuations are uncorrelated, as in equilibrium, $\alpha=1/2$, whereas suspensions 
with  correlated density fluctuations will be characterized by an exponent $\alpha > 1/2$.  
\textcolor{black}{The system is characterized by  giant density fluctuations (GDF) when $\alpha \geq 1/2$ ~\cite{ramaswamySC, chatePRE}. }
Active systems have been shown to display large number density fluctuations, as
  in bacterial colonies~\cite{swinney} with  
  a scaling exponent  $\alpha = 3/4 \pm 0.03$, or 
   in computer models of self propelled particles~\cite{fily,chatePRE}
   with   a maximum value $\alpha =0.8$. \textcolor{black}{These fluctuations in active systems can have different nature: Vicsek-like models, require  long-range orientational order~\cite{chatePRE} whereas the fluctuations of self-propelled rods (SPR) are remarkably different \cite{weitzSPR} since either polar or apolar long-range or quasi-long-range order are absent. In our case the hydrodynamic signature is what leads to polar order  and the subsequent emergence of fluctuations. \cite{weitzSPR,Dey}}
  
\begin{figure}
\begin{center}
\includegraphics[clip,scale=0.8]{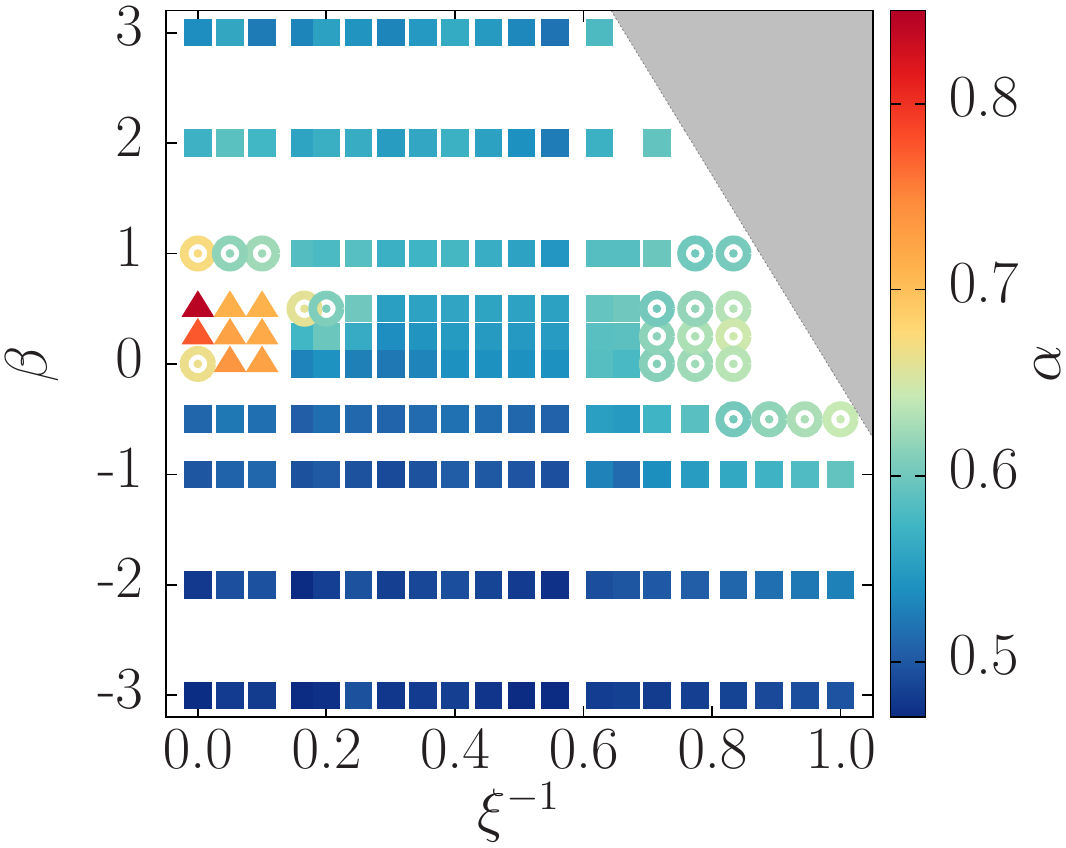}
\caption{Colour map of the scaling exponent $\alpha$ for values of $\xi$ and $\beta$ where coarsening is absent. 
Triangles correspond to $\alpha > 0.7$, circles to $\alpha$ between $0.6$ and $0.7$, where partially large fluctuations are
 found,  and blue squares are for homogeneous suspensions where $\alpha < 0.6$. The gray area represents the values where coarsening is found.}
\label{exponent_fluc}
\end{center}
\end{figure}
  
In squirmer suspensions we find that $\alpha$ depends both on  active stresses and attractive 
interactions, as shown in Fig.~\ref{exponent_fluc}  
(where the grey area indicates the region of the phase space 
where coarsening is observed).  
 We estimate exponents between ($0.47,0.85$) and identify three regimes: one with scaling exponent of
  $\sim 0.5$ for homogeneous suspensions (blue squares), a second one with intermediate values between
   $0.6$ and $0.7$ (green-yellow circles) and a third one when suspensions present large and loosely 
   packed dynamic clusters $\alpha > 0.7$, as in the case of squirmers with 
   $\xi > 5.0$ and $0 < \beta < 1$ (red-orange triangles).

 For the same degree of active stresses, 
   pushers and pullers  do not always exhibit the same  number density fluctuations: 
       while for pushers  
$\alpha$ remains close to $1/2$  and grows very slowly as the attraction strength increases 
   ($\xi^{-1} \rightarrow 1 $), reaching values of $\alpha = 0.64$ for $\beta=-0.5$ and $\xi=1$;  pullers with $\beta > 1$
    have an exponent around $0.6$, independently of the attraction strength. A value of $0\leq \beta \leq 1$ leads to 
    anomalous density fluctuations  due to     stress activity 
from a maximum value of $\alpha$ at $\xi = \infty$ to a value typical for homogeneous suspensions 
($\alpha \rightarrow 0.5$) as $\xi \rightarrow 1$. 

To conclude, we observe  GDF for pullers and for pushers with small $|\beta|$, stressing the relevance 
of hydrodynamic coupling  in the morphology of the cluster phase.   \textcolor{black}{We observe that $\alpha>3/4$, for weak pullers $(0< \beta < 1)$, correlate with the development of large clusters. Therefore, GDF with large values of $\alpha$ constitute  a signal of collective morphological changes.}

\subsection{Cluster-size distribution}

In order to quantify how the interplay between attractive forces and activity determine the dynamic clusters, 
we have systematically analysed  the clusters structures  when semidilute active suspensions reach a
 steady state, by  studying the cluster size distribution (CSD) as a function of the activity parameter $\xi$, and 
 the active stress generation $\beta$ (for squirmers).
 
  \begin{figure}[h!]
\begin{center}
\includegraphics[clip,scale=0.6]{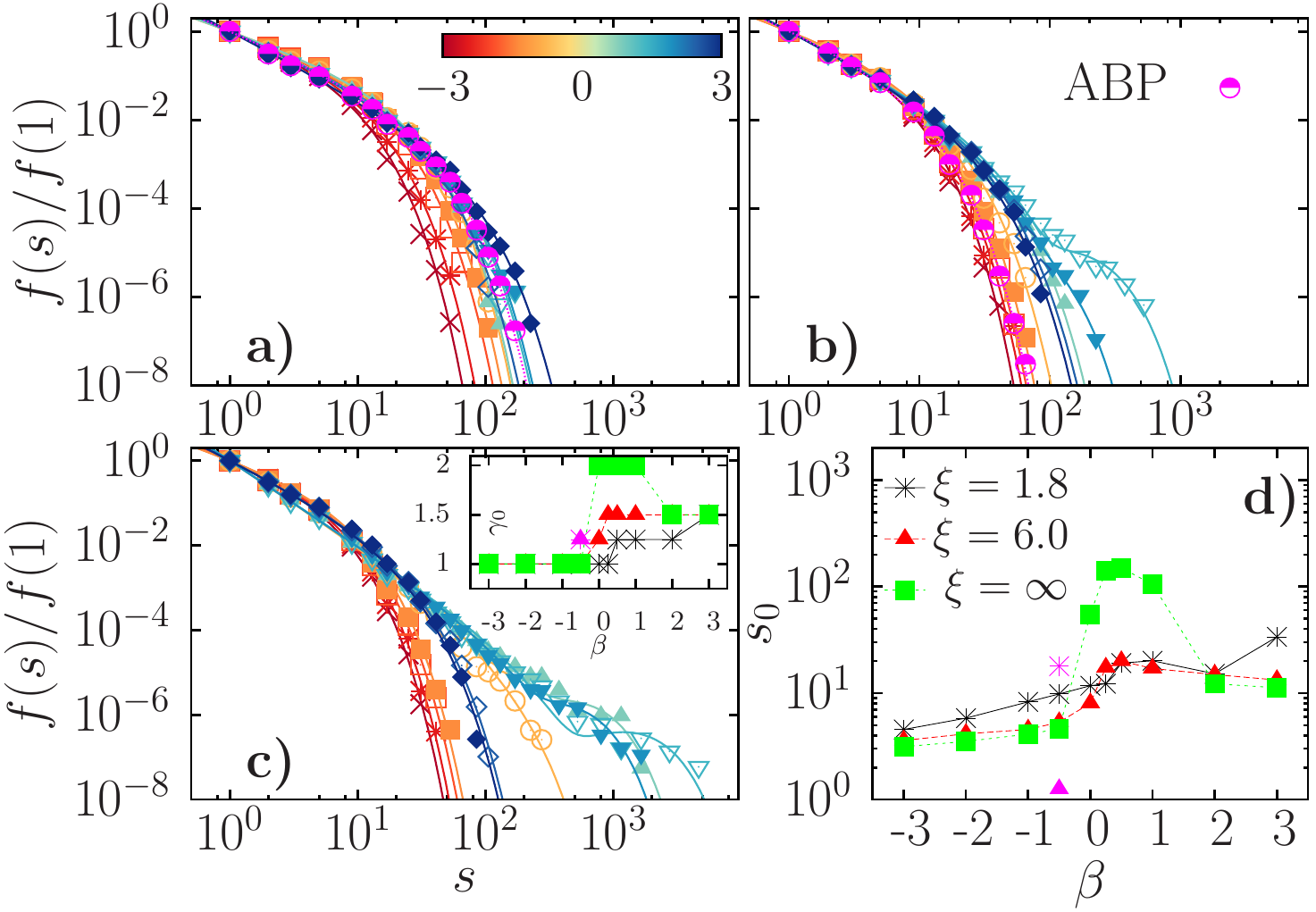}  
\caption{Cluster-size distribution $f(s)$ for different values of $\xi$  and $\beta$. Red symbols correspond to 
pushers $\beta < 0$ and blue ones to pullers $\beta > 0$, yellow for $\beta=0$. Pink points 
represent the CSD for ABP. Solid lines correspond to fitted curves of eq. (\ref{eq:csd_cutoff}). 
a) CSDs with $\xi=1.8$, b) CSDs with $\xi=6.03$, c) CSDs for squirmers with $\xi=\infty$, the inset  showing the fitted 
values of the exponent $\gamma_0$ as a function of $\beta$ for three different interaction strengths and in pink the fitted parameter for ABP. 
d) Cluster size cut-off $s_0$ as a function of $\beta$ for three different values of $\xi$. $s_0$  obtained by fitting $f(s)$ to equation (\ref{eq:csd_cutoff}), in pink the fitted parameter for ABP.}
\label{f_s}
\end{center}
\end{figure}

  Figure~\ref{f_s} displays the
  CSD for different values of  $\xi$, for $\beta$ in the range between -3 and 3, calculated as the average fraction of 
  clusters $f(s)$ of a given number of particles $s$. Fig. \ref{f_s}-a shows CSD for $\xi=1.8$: the CSD 
  are wider as $\beta$ increases without reaching a purely algebraic function ($f(s)\sim s^n$). ABP  with $\xi=1.8$  are characterized 
  by   CSD resembling those of weak pushers. 
  
For $\xi=6.03$ (Fig. \ref{f_s}-b) CSD for pushers ($\beta < 0$) shows a very weak dependence on
 the magnitude of the active stress and have approximately the same shape. For pullers CSDs are 
 wider, and pullers with small $\beta$ develop an inflection point in the distribution for large 
 clusters ($s\sim 10^2$). As $\beta$ increases to $\beta > 1$, the CSDs narrows  even further. 
 Whereas ABP display a CSD that   resembles those observed for weak pushers, 
 even though for smaller clusters.

In the case of repulsive squirmers   (Fig. \ref{f_s}-c, $\xi=\infty$), the CSDs for pushers are 
narrower and have a very weak dependence on $\beta$ as in the case of pushers with $\xi=6.03$ 
(red curves in Fig. \ref{f_s}-c); whereas for weak pullers, $\beta\simeq 1$, the distributions are wider 
and characterized by an inflection point more marked at $s \approx 10^3$ for $\beta=0.5$; for $\beta>1$ the 
distributions get narrower again but not as much as the corresponding ones for pushers. 
 It is important to notice that, even in the 
absence of any attraction,  generically  the CSD width increases only with the active stress (through $\beta$).
When $\beta=0$ the CSD starts showing  
an inflection point. 
 For ABP without any attractive 
 interaction we shall expect a very narrow CSD since both density and activity are too small to allow for  cluster formation. 

From the panels a, b and c of Fig. \ref{f_s} we deduce that attraction diminishes the probability 
to generate an inflection point in the CSD for $0< \beta < 1$. This feature indicates that  the  dynamics of large clusters has fundamentally an 
hydrodynamic origin.

The   CSD can be accurately fitted by
\begin{equation}\label{eq:csd_cutoff}
\frac{f(s)}{f(1)}= A~\frac{\exp(-(s-1)/s_0)}{s^{\gamma_0}}+B~\frac{\exp(-(s-1)/z_0)}{s^{-\gamma_0}},
\end{equation}
with $\gamma_0$, $s_0$, $z_0$ and $B$ constants that depends on $\beta$ and $\xi$, such that $A=1-B$. 
The parameter $s_0$ has been used to control the cutoff at large cluster size~\cite{demian} 
and  related  with the  particles density \cite{peruani06,peruani13}.
More recently, the divergence of the cluster size cut-off has been related to the location 
of the phase separation in a suspension of Brownian self-propelled repulsive disks \cite{demian,fily}.

Since in Fig.~\ref{f_s}-a, for  $\xi=1.8$ we do not observe   an inflection point,  we take  $B=0$  and  $\gamma_0=1$ for pushers and
  pullers with $\beta < 1/2$. The value of $\gamma_0$ will then grow as a function of $\beta$ (black
   points in the inset  of Fig. \ref{f_s}-c). 
   The CSD for ABP with $\xi=1.8$ can be fitted with 
   $s_0=17.9$  and the exponent $\gamma_0=1.25$:  which are values close to the ones 
for squirmers with $\beta$ slightly negative.

The CSDs for pushers with $\xi=6.03$ corresponds to $B=0$ (no inflection),   
an exponent $\gamma_0=1$, and $s_0$  almost independent of $\beta$, since the distributions are pretty similar. 
The CSD for ABP is fitted with $s_0=5.14$ 
and $\gamma_0=1.25$,  parameters that are similar to the ones fitted for the CSD of pushers with $\beta$ between $-1$ and $-0.5$.

Once $\beta$ is non-negative, $\gamma_0$ 
grows as a function of $\beta$ as in the previous case with larger attractions between 
particles (red circles in the inset  of Fig. \ref{f_s}-c), but in this case $B =4\times10^{-8}$ 
and $7\times10^{-7}$ for $\beta=0.5$ and $1$, respectively, due to the presence of an inflection  in the 
distributions (and $B=0$ otherwise).

The CSDs for pushers with $\xi=\infty$ also present $B=0$ and $\gamma_0=1$ like pushers
 with $\xi=6.03$. On the other hand, when $0\leq \beta \leq 1$ the CSDs are fitted with $B \neq 0$ where $10^{-12}\leq B \leq 10^{-9}$ and $B=0$ for $\beta > 1$.

To summarize, the CSD curves for pushers never develop an 
inflection point independently of the interaction strength, thus $B=0$, 
 the power law exponent never change and $\gamma_0=1$ in all cases. The transition to 
 another value of the exponent $\gamma_0$ depends on the interaction strength, as shown 
  in the inset  of Fig. \ref{f_s}, where $\gamma_0$ changes with  $\beta$ depending on the interaction strength.

 Figure \ref{f_s}-d shows how the characteristic cutoff  cluster size $s_0$, increases with 
 $\beta$ when  the attractive interactions compete with activity ($\xi=1.8$). The general
  increase when we move from pushers to pullers always persists, indicating  that pullers
   favours the development of larger cluster sizes. However, when attractive forces become 
   sub-dominant with respect to self propulsion, the region of weak pullers (when self propulsion 
   and active stresses are comparable) is characterized by much wider CSDs which are reflected
    in a strong increase in the dependence of $s_0$ on $\beta$, with a dominant peak at 
    $\beta = 1/2$. Finally, in all the cases where $B \neq 0$, we found that $z_0$ has values similar (the same order of magnitude) to $s_0$.

In the case of the CSDs for ABP we found the exponent $\gamma_0=1.25$ for all the
 studied interaction strengths, whereas $s_0$ was growing as the attraction increased.

\subsection{Radius of gyration}

\begin{figure}[h!]
\begin{center}
\includegraphics[clip,scale=0.575]{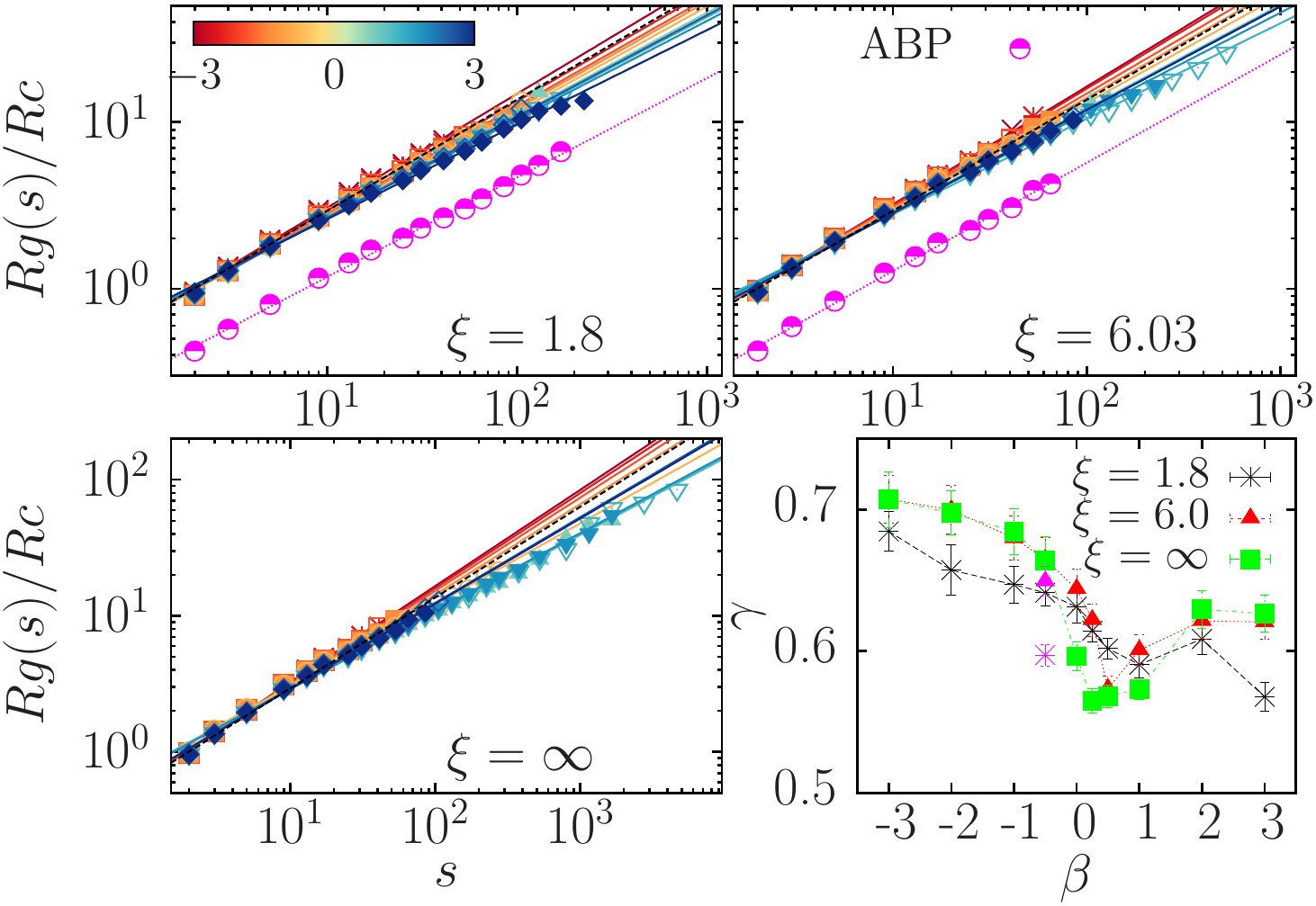}  
\caption{Radius of gyration $R_g(s)$ (normalized by the particle's radius) 
as a function of the cluster-sizes for values of $\xi$ ranging from 1.8 up to $\infty$ and $\beta$ between $3$ and $-3$. 
Red symbols correspond to pushers $\beta < 0$ and blue ones to pullers $\beta > 0$. Solid lines represent $Rg(s) \sim s^{\gamma}$. 
Top-left $\xi=1.8$,  and top-right: $\xi=6.03$. 
Pink data correspond to $R_g$ for ABP with the same values of $\xi$. 
Bottom-left: $\xi=\infty$ and Bottom-right: $\gamma$ exponent as a function of $\beta$ and $\xi$.}\label{RG}
\end{center}
\end{figure}

In order to characterize the cluster morphology, we have computed the dependence of the clusters  radius of 
  gyration $Rg(s)$ with their size, $s$,  from Eq.~(\ref{eq:R_g}).  Fig.~\ref{RG} displays the mean 
  radius of gyration normalized by the particle radius. 

  We always observe an algebraic dependence, 
  $Rg \sim s^{\gamma}$, with an exponent that depends both on $\xi$ and $\beta$.

   When $\xi=6.03$ (top-right  in fig.~\ref{RG}), $Rg(s)$ follows an exponent greater than $2/3$ for pushers, 
whereas $R_g(s)$ decreases slower for pullers than for pushers.
 In fact the exponent $\gamma$  depends on the $\beta$ value:  
 the larger the   $\beta$ the larger the exponent, even though always smaller than $2/3$.
 The $R_g(s)$ slope decreases with $\beta$ up to $\beta=0.5$ where the minimum 
  value for $\gamma$ is reached, then $\gamma$ increases as $\beta$ increases.

  Fig.~\ref{RG} 
  (bottom-right) displays the dependence of the scaling exponent on $\beta$ for different values of $\xi$. 
  For pushers the scaling exponent decreases as $\beta$ increases. This tendency persists initially for
   pullers, but changes its trend above $\beta\simeq 0.5$, leading to cluster re-expansion, and increases 
   afterwards to saturate around $\gamma \simeq 0.62$. Puller clusters are more compact than their 
   pusher counterparts. 
      When activity  dominates the cluster morphology becomes indistinguishable from 
   that of repulsive squirmers. 
   In fact, comparing the curves for $\xi=1.8$ and $\xi=\infty$ one can 
   see that  active hydrodynamic stresses determine the main changes of the scaling exponent 
   independently of the attractive forces. In particular, the re-expansion observed for pullers increasing 
   active stresses is essentially a hydrodynamic phenomenon. As the attractive forces increase their
    relative importance, the clusters become more compact but the dependence on $\beta$ is not significantly altered.

     In the case without hydrodynamics, clusters are more compacted as we increase the 
  interaction strength,  as expected. 
  $R_g$ of ABP clusters  follow a power law with an exponent of $\gamma=0.6$ for $\xi=1.8$ 
which is similar to the exponent for squirmers with slightly negative $\beta$
at the same interaction strength; if $\xi=6.03$
 the $R_g$ for ABP develops an exponent of $\gamma=0.65$ which corresponds to the exponent developed 
 by squirmers with a slightly negative $\beta$ with the same interaction strength.

The values of $\gamma$ falls between  the scaling exponent corresponding to diffusion limited cluster 
aggregation (DLCA), 0.704, and the reaction limited cluster aggregation (RLCA), 0.62. 
As we move from pushers to 
pullers we see that clusters become more compact and get closer to RCLA. 
The clusters that active suspensions form are always more compact than DLA ones, with $\gamma = 0.588$, 
and definitely more compact than the structure of percolating clusters, with $\gamma= 0.527$. Only for $\beta =0.5$, squirmers 
have an exponent close to the DLA one. This trend is consistent with the  increase in  density fluctuations and the wider 
CSD observed for weak pushers. 

However, the connection between physical interactions and cluster aggregation mechanisms 
remains elusive. Even though in colloidal systems the fractal structure of colloidal clusters is thought to be very sensitive 
to the nature and range of  particle attractions~\cite{roque}, we conclude that the crossover from RLCAS to DLCAS-like behavior 
 for squirmers is  essentially a hydrodynamic phenomenon.

\subsection{Polar order}
\begin{figure}[h!]
\begin{flushleft}
\includegraphics[clip,scale=0.575]{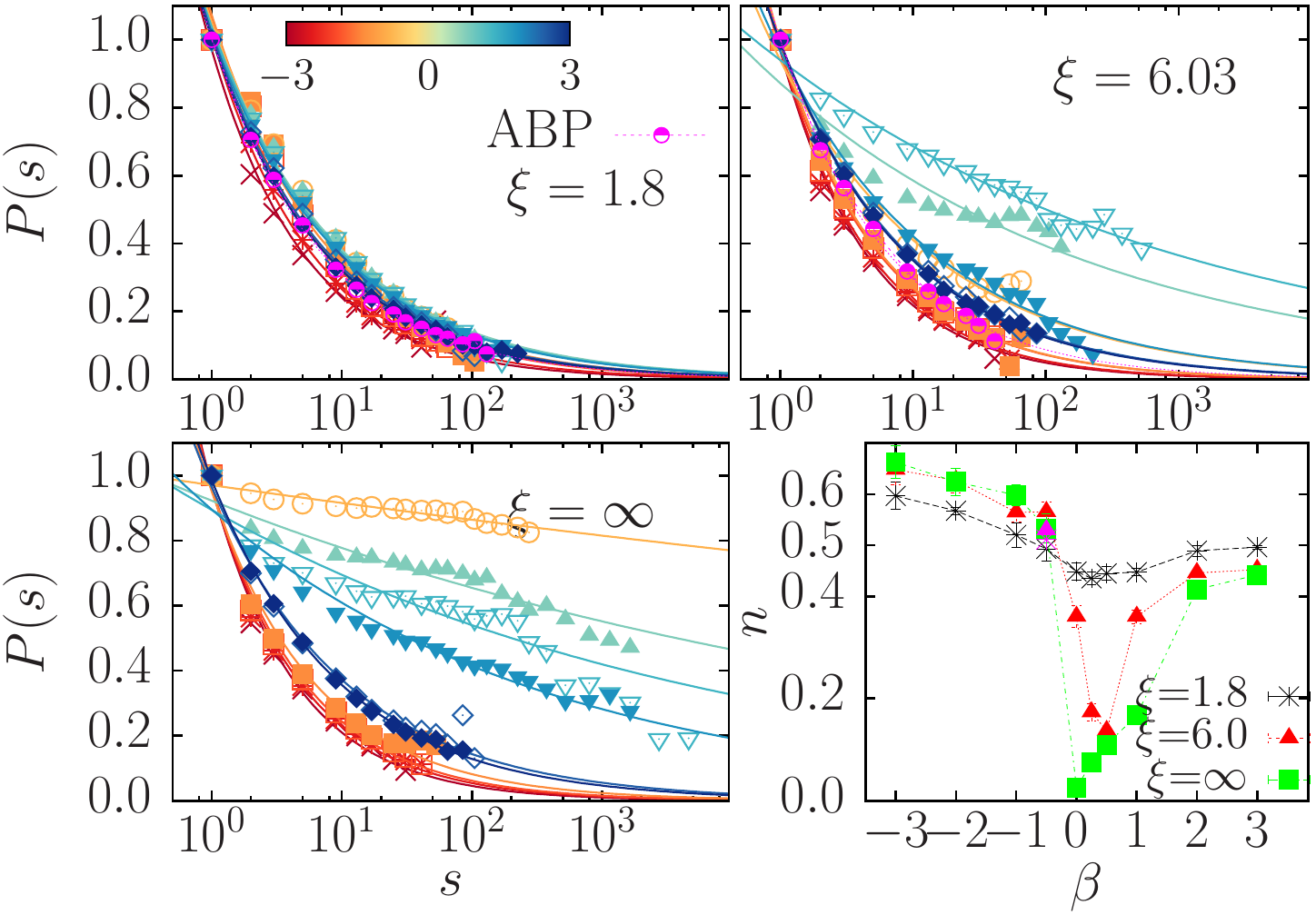}  
\caption{Polar order parameter $P$ 
as a function of the cluster-size for values of $\xi$ ranging from 1.8 up to $\infty$ and $\beta$ between $-3$ and $3$. 
Red symbols correspond to pushers $\beta < 0$, blue ones to pullers $\beta > 0$, yellow  to $\beta=0$. Solid lines correspond to the fitted curve
 $P(s) \sim s^{-n}$ and pink circles  to ABP. Bottom-right: Exponent $n$ as a function of $\beta$ and $\xi$, in pink the fitted parameter for ABP.}
\label{polar}
\end{flushleft}
\end{figure}

We analyse the degree of alignment as a function of the cluster size, using Eq.~ (\ref{eq:P}). 
Fig.~\ref{polar} shows that, generically, the degree of polar order decreases with cluster size. 

Pushers (in red) have a faster decay than pullers (in blue). In the absence
 of significant attraction, for $\xi=\infty$, we see that for pushers the order decreases algebraically, 
 with a dependence compatible with $P(s)\sim s^{-1/2}$, while pullers show a persistent polar order 
 for all  cluster sizes. In particular, for $\beta=0$ we observe  a net polar alignment for all clusters. 
 This strong polar ordering is consistent with the development of  a global polar phase that displays 
 its maximum orientation in $\beta=0$ \cite{alarcon2013}.

 As the attraction plays a more relevant role, the degree of polarity  for pullers  decays and eventually 
  the degree of ordering in the clusters  is  independent of $\beta$. 
    Therefore, the decay $P(s)\sim s^{-1/2}$ 
  is essentially due to the competition between  self propulsion and attraction
  even though near field  hydrodynamic coupling plays a relevant role 
    because in all cases, even at small $\xi$, ABP clusters show a larger degree of polar ordering than pusher squirmers.
For ABP, at $\xi=6.03$ and $\xi=1.8$, $P(s)$ has a slower decay as $s^{-1/2}$. 
This means that particles within a cluster maintain their polar order better than  pushers do, 
due to the fact that the rotation induced by hydrodynamics helps to destroy polar ordering~\cite{ricard_softmat2015}. 

When $\xi =6.03$ (top-right panel) pushers show a decay with an exponent $n>0.5$, whereas pullers with $\beta <1$ 
 drastically change their behaviour compared to pullers at $\xi=1.8$ showing big clusters with high 
 polar order. Clusters of ABP on the contrary, do not suffer any change in their polarity 
 and $P(s) \sim s^ {-1/2}$ independently of the interaction strength.

\begin{figure}[h!]
\begin{center}
\includegraphics[clip,scale=0.575]{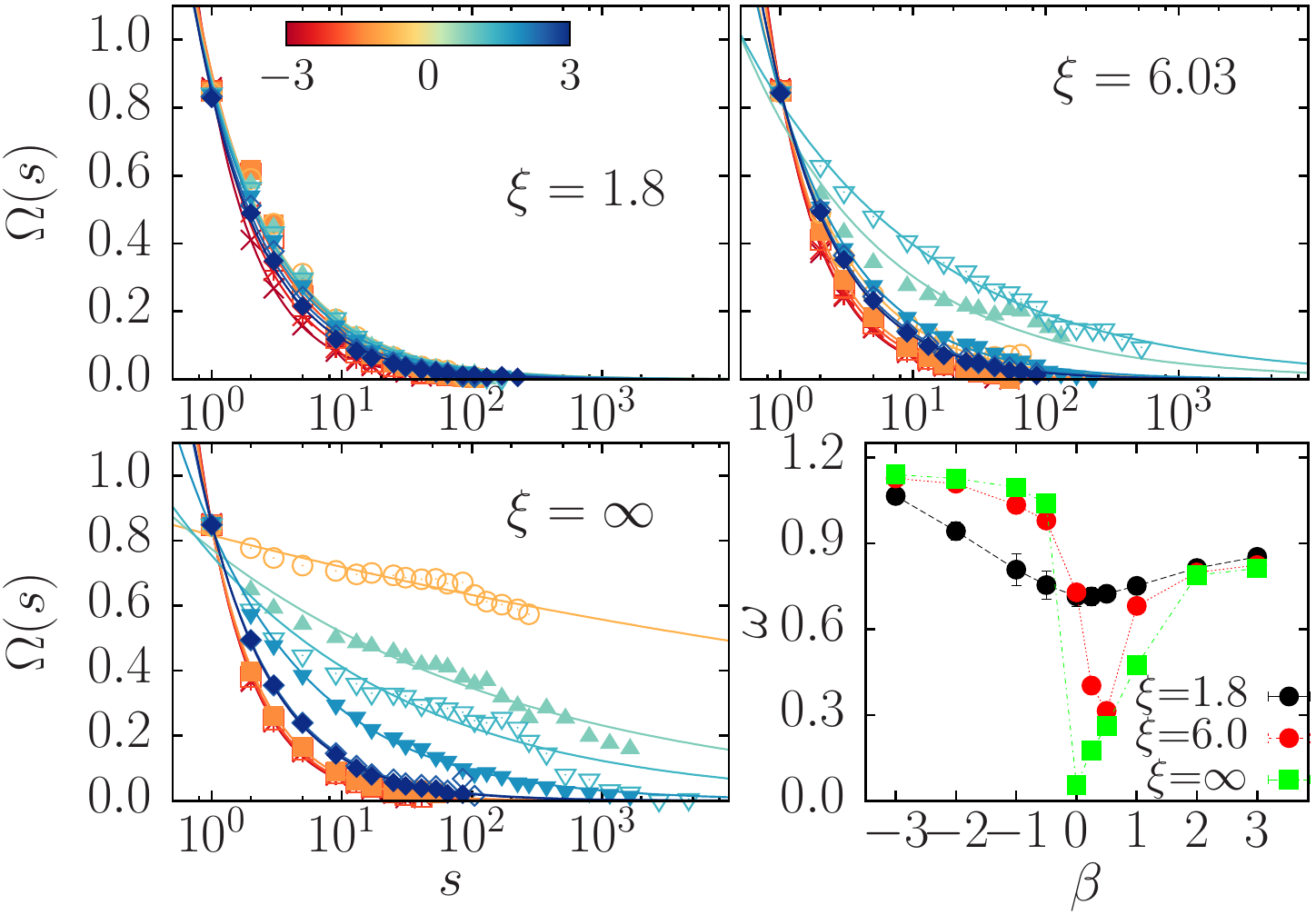}  
\caption{$\Omega$ as a function of the cluster-size for values of $\xi$ ranging from 1.8 up to $\infty$ and $\beta$ 
between $-3$ and $3$. Solid lines correspond to $\Omega(s) \sim s^{-\omega}$ Top-left: $\xi=1.8$, Top-right: $\xi=6.03$, 
Bottom-left:  $\xi=\infty$. Bottom-right: $\omega$ exponent value fitted as a function of $\beta$ and $\xi$. $\Omega(s)$ is 
normalized by the overdamped velocity reached by one squirmer ($v_s= 2/3 B_1$).}
\label{omegapolar}
\end{center}
\end{figure}

In order to clarify  whether  clusters move in the same direction as the particles that constitute them, 
we have computed for the squirmers case the cluster orientation with respect to the cluster's center-of-mass 
velocity, $\Omega$, as a function of the cluster size, according to Eq.~(\ref{eq:Omega}), as shown in Fig.~\ref{omegapolar}.  

In all cases we observe a decay of $\Omega$ with cluster size, indicating that the direction of motion of the cluster 
is progressively decorrelated from the average cluster alignment. The slower decay in $\Omega$ at large $\xi$
 for pullers, when hydrodynamic stresses dominate, indicates that there is a strong correlation between
  the degree of  polarity and the direction of motion of the cluster. 

When attraction is reduced (top-right panel Fig.~\ref{omegapolar}) pullers with $\beta < 1$ have a slow decay given 
that it is possible to find a partial orientation for big clusters (as shown in figure \ref{polar}). When $\beta = 0$ without 
attractions the clusters orientation decay is very slow (yellow circles in the bottom-left panel of Fig. \ref{omegapolar}), 
almost independent of cluster size, given that in this case the global aligned state is stable. Therefore,  we have fitted 
 $\Omega(s)$ to a power law $s^{-\omega}$, and the corresponding curves are displayed in Fig.~\ref{omegapolar} 
 (bottom-right panel), showing a reasonable agreement. The bottom right panel of Fig.~\ref{omegapolar}) shows the dependence of $\omega$ both 
  on $\beta$ and $\xi$. In general, the decay of $\Omega$ is slower for pullers than for pushers, due to the larger degree 
  of polar order of the former. As $\xi$ decreases, the attractive interactions become more dominant and the dependence 
  on active stresses weakens. However, even for the smallest $\xi$ we still see that pushers  have a weaker degree of correlation.

\section{Discussion and conclusions}\label{Sec:conclusions}
	
In this paper we have carried out  a systematic study of semi-dilute suspensions of interactive squirmers restricted to swim 
in a plane surrounded by  an unconfined fluid. We have seen that active hydrodynamic stresses give rise to  a steady state
 characterized by  dynamic  clusters. The properties and character of these clusters are sensitive to the  competition between 
 self propulsion, induced hydrodynamic flows through active stresses and direct  short-ranged attractive Lennard-Jones
  particle isotropic interactions. We have identified the relevant dimensionless parameters that determine the competition 
  between self-propelling and attractive forces, $\xi$, which can be regarded as an effective P\'eclet number, and the relative relevance between self propulsion and active stresses, 
  $\beta$. 
  
  The systematic study of the system as a function of  ($\xi$, $\beta$), has allowed us to identify three regimes associated to the intrinsic motion of such suspensions. When squirmers attractions dominate over self-propulsion, squirmers aggregate and display coarsening. When attractive forces competes with self propulsion a steady state is reached, where squirmers  are characterized by a cluster size distribution. The transition between these two regimes is a function both of $\xi$, and the character of the induced active stresses, $\beta$. This fact already shows that the  collective behaviour of an active suspension is not only a  function of its degree of activity, quantified by $\xi$, but rather on the details of the hydrodynamic coupling. When activity dominates we have seen that in some situations the steady state sustains an oscillatory behaviour. In this third regime we have analysed the collective behavior over time scales large compared to these intrinsic oscillations.

      We have focused on the regimes 
   where  dynamic clusters are stable, typically $\xi >1$,  and have quantified the 
   cluster size distributions  to classify the  emerging steady states. Generically, all the CSDs can be fitted to a bimodal 
   cutoff algebraic shape.  The shape of this CSDs is basically prescribed by hydrodynamic interactions. The Lennard-Jones attraction modulates the  cutoff distance and leads typically to larger clusters for a given value of 
$\beta$, except in the regime of weak pullers, where attraction  disrupts significantly the bimodal shape of the CSD 
and the strong density fluctuations that characterize this regime. In this case attraction increases significantly the cutoff 
distance for the CSDs. 
\textcolor{black}{ These fluctuations in the mean size aggregate and the bimodal shape of the CSD emerge purely because of hydrodynamic interactions, since they appear only for weak pullers and very low attraction. Moreover, the active stress given by $\beta$ re-orients and polarizes the particles in the dynamic clusters (see Figs. \ref{polar} and \ref{omegapolar} and movies in the Supplementary Information). A similar coordinated behaviour has been observed with SPR without hydrodynamic interactions \cite{weitzSPR}, where aggregates with intrinsic topological defects  grow. Such defects  originate  highly compressed  structures, whose   active stresses   lead to the  fluctuations in the aggregate boundary and the ejection of polar clusters. Thus, the analogy between both systems suggests that aggregate fluctuations might be not model-dependent. One could try to map the active stress induced  by $\beta$ in squirmer aggregates   with the active stress induced by  topological defects in SPR. Alternatively, one could  tune  the squirmer aspect ratio~\cite{ishikawaPRE15} since it is known that elongated squirmer suspensions develop a variety of collective motion, including  ordering, aggregation and whirls; such morphologies depend on the hydrodynamic signature, given by $\beta$, and the particle shape.}

\textcolor{black}{In this work we have considered the general case of disks and spheres to spotlight purely the effect of the hydrodynamic interactions. We believe that particle shape plays an important role in the formation of emerging collective patterns \cite{ishikawaPRE15}, and it deserves a study on its own.}
  
  We have further compared the properties of the cluster phase  of squirmers with the one generated by Active Brownian Disks. In the latter case 
 there are no stresses due to the surrounding fluid, and the only relevant parameter is $\xi$, the effective P\'eclet number.
We have found that when attraction competes with activity, ABP form larger clusters than squirmers at equivalent $\xi$ and the properties of the cluster phase 
resemble those of weak pushers.
 However, when activity dominates the CSDs are similar to the CSDs of pushers with the same interaction strength. 
 
The radius of gyration of the observed clusters also identifies the impact of active stresses. We have found that the 
exponent of the dependence of the radius of gyration on cluster size corresponds to clusters that are always more
 compact than DLA and move from DLCA to RLCA as active stresses change from pushers to pullers. Attractive 
 forces lead to more compact clusters, but do not change qualitatively the trend promoted by active stresses. When 
 attractive interactions compete with activity we find that clusters shrink as $\beta$ grows, while  when activity
  dominates a contraction-expansion effect occurs for pullers. Squirmers form   generically polarized clusters, but the degree 
  of ordering decreases with cluster size, showing again a strong difference between pullers and pushers.The polar order of 
  the clusters is reduced when hydrodynamic interactions are present in the case where attraction competes with the activity, 
  thus ABP clusters are more aligned. This characteristic effect however is not observed once the activity dominates the attractive interactions.
    
Therefore, we have shown that hydrodynamics alone can sustain a cluster phase of active swimmers (pullers), while ABP form cluster 
phases due to the competition between activity and self propulsion. The structural properties of the cluster phases of squirmers
 and ABP are similar, although squirmers show  sensitivity to active stresses. ABP resemble weakly pushers squirmer 
 suspensions in terms of CSDs, structure of the radius of gyration on cluster size and degree of cluster polarity.

\section*{Acknowledgements}
FA acknowledges  CONACYT-CECTI (M\'exico)
through Grant No. 214151 for financial support. FA and IP thank  MINECO (Spain) through FIS2015-67837-P, DURSI Project 2014SGR-922. IP also acknowledges the support of Generalitat de Catalunya under program Icrea Academia for financial support.
CV acknowledges the EU project 322326-COSAAC-FP7-PEOPLE-CIG-2012, the National Project FIS2013-43209-P and a Ramon y Cajal tenure track.
We thank the KITP at the University
of California, Santa Barbara, where they were supported
through National Science Foundation Grant NSF PHY11-25925.
This work was possible thanks to the access to MareNostrum Supercomputer at Barcelona Supercomputing Center (BSC) and also through the Partnership for Advanced Computing in Europe (PRACE). This article is based upon work from COST Action MP1305, supported by COST (European Cooperation in Science and Technology)

\providecommand*{\mcitethebibliography}{\thebibliography}
\csname @ifundefined\endcsname{endmcitethebibliography}
{\let\endmcitethebibliography\endthebibliography}{}



\bibliographystyle{rsc} 

\clearpage 

\section{Appendix}
\subsection{Squirmer model}
\label{appx:squirmermodel}

We have used the well known model for  microswimmers developed by Lighthill \cite{Lighthill} and improved by Blake \cite{Blake} (proposed to model the locomotion of ciliated microorganisms) in order to simulate self-propelled particles with hydrodynamics interactions. 
 This mode, named squirmer, considers a spherical particle with an internal   activity that induces an effective axisymmetric velocity on its surface. Accordingly,  the velocity field id defined in terms of  two components that depend on the two polar coordinates $(r,\theta)$. Therefore, this velocity can be written as two independent terms, a radial $u_r$ and a polar $u_{\theta}$ term

\begin{align}\label{slip_veloc}
u_{r}|_{r=R_p} &= \sum_{n=0}^\infty A_n\left(t\right)P_n\left(\frac{\textbf{e}_1 \cdot\textbf{r}}{R_p}\right), \nonumber \\
u_\theta|_{r=R_p} &= \sum_{n=0}^\infty B_n\left(t\right)V_n\left(\frac{\textbf{e}_1 \cdot  \textbf{r}}{R_p}\right),
\end{align}
\noindent where $\textbf{r}$ represents the position vector  with respect to the squirmer's center (always pointing to the particle surface so that $|\textbf{r} | = R_p$); ${\bf e}_1$  prescribes the intrinsic self-propelling direction, (which moves rigidly with the particle and determines the direction along which a single squirmer displaces) ; $P_n$ stands for the $n$-th order Legendre polynomial and $V_n$ is defined as
 
\begin{eqnarray}
V_n\left(\cos \theta \right)= \frac{2}{n(n+1)}\sin \theta~ P'_n(\cos \theta).
\end{eqnarray}

The amplitudes $A_n(t)$ and $B_n(t)$ are periodic functions determining the flow induced by the beating cilia on the squirmer's surface and $\theta$ is the angular polar coordinate. Since the cilia wave stroke is faster than the squirmer displacement, we can replace the time dependent
amplitudes at the boundary, eqn. (\ref{slip_veloc}), by their effective averaged amplitudes over a stroke period, $B_n(t) = B_n$. Moreover, we neglect the radial changes of the squirming motion, $A_n(t)=0$. 

At low Reynolds numbers, the flow induced by a squirmer can be obtained by solving the Stokes and continuity equations
\begin{align}\label{Stokes}
\nabla p &= \nu \nabla^2 \textbf{u},
 \nonumber \\
\nabla \cdot \textbf{u} &= 0.
\end{align}
for the fluid  pressure, $p$, and velocity fields, $\bf{u}$, subject to the boundary conditions, Eq.~\ref{slip_veloc}~\cite{Ishikawa,Llopis2013}. The  mean fluid flow induced by squirmer can be expressed as
 \begin{align}\label{gralvel}
 &\textbf{u}\left( \textbf{r}\right) = B_1\textbf{e}_1 \left( -\frac{1}{3} \textbf{I}+ \frac{\textbf{r} \textbf{r}}{r^2} \right) \left( \frac{R_p}{r} \right)^3 \nonumber \\
 &+
 \sum_{n=2}^\infty B_n \left( \frac{R_{p}^{(n+2)}}{r^{(n+2)}} - \frac{R_{p}^n}{r^n} \right) 
 P_n\left( \frac{\textbf{e}_1 \textbf{r}}{r} \right) \frac{\textbf{r}}{r}  \nonumber \\
 &+
  \sum_{n=2}^\infty B_n \textbf{e}_1 \left( 1 - \frac{\textbf{r} \textbf{r}}{r^2}
  \right) \nonumber \\
  &\times \left( \frac{n R_{p}^{(n+2)}}{2 r^{(n+2)}}
  - 
  \frac{(n-2) R_{p}^n}{2 r^n} \right) \frac{V_n \left(  \frac{\textbf{e}_1 \textbf{r}}{r}\right)}{\sqrt{1 - \frac{\textbf{e}_1 \textbf{r}}{r}}}.
 \end{align}
A squirmer particle swims force and torque free at a constant speed of magnitude $v_s = \frac{2}{3}B_1$ along $\textbf{e}_1$ with respect to the solvent. 

We consider a simplified version of the squirmer model, and  take $B_n = 0$ when $n > 2$, keeping only the first two terms in the general expression of Eq.~(\ref{gralvel}). The two non-vanishing terms are enough to model two essential features of the impact that squirmers have on the surrounding medium: while the polarity is related to the squirmer self - propulsion through the coefficient $B_1$,  the active stresses are induced by the apolar term $B_2$. The squirmer's active stress can be quantified in terms of the squirmer self-propulsion by $\beta = B_2/B_1$ \cite{Ishikawa}: if $\beta > 0$ the squirmer behaves as puller whereas if $\beta < 0$ it behaves as a pusher. Therefore, the average fluid flow generated in this simplified model by a squirmer can be written as

\begin{align}\label{fluidflow}
 \textbf{u} \left( \textbf{r} \right) &= -\frac{1}{3} \frac{R_{p}^3}{r^3} B_1\textbf{e}_1 + B_1\frac{R_{p}^3}{r^3} \textbf{e}_1 \cdot \frac{\textbf{rr} }{r^2} \nonumber \\
   &- \frac{R_{p}^2}{r^2} B_2 P_2 \left(\frac{\textbf{e}_1 \textbf{r}}{r} \right)\frac{\textbf{r} }{r}.
 \end{align}

The first two terms of equation (\ref{fluidflow}) represent a dipolar field, similar to the one generated by an electric dipole. The direction and strength of the fluid flow is specified by the polarity term $B_1 {\bf e}_1$ in analogy with the electric moments. In turn,  the $B_2$ term models a quadrupolar field. $B_2$ is equivalent to the strength of a quadrupole for a symmetric arrangement of electric dipoles, when the dipole moments vanish . Then, taking into account that we have only two non-zero terms, the boundary conditions on the surface of the squirmers depicted in equations (\ref{slip_veloc}) can be written as
\begin{align}\label{slip_veloc_bn0}
u_r|_{r=R_p} &= 0, \nonumber \\
u_\theta|_{r=R_p} &= B_1 V_1(\cos \theta) + B_2 V_2(\cos \theta).
\end{align}

Where $u_r$ and $u_{\theta}$ are the radial and tangential components of the fluid velocity 
$\textbf{u}$.

\subsection{Identifying clusters in the suspension. \label{sec:clusters}}

To start with, we compute the radial distribution function  of the Lennard-Jones fluid at the $\phi$ studied in the manuscript (only considering configurations in the time interval when the system is in steady state).
From Figure 9 we  already observe that the system with pullers is more ordered than the system with pushers. 
In order to detect the clusters in the suspension, we use the first minimum ($r_{cl}$) of the radial distribution function and identify particles within this distance as neighbours belonging to the same cluster.

\begin{figure}[h!]
\begin{flushleft}
\includegraphics[clip,scale=0.625]{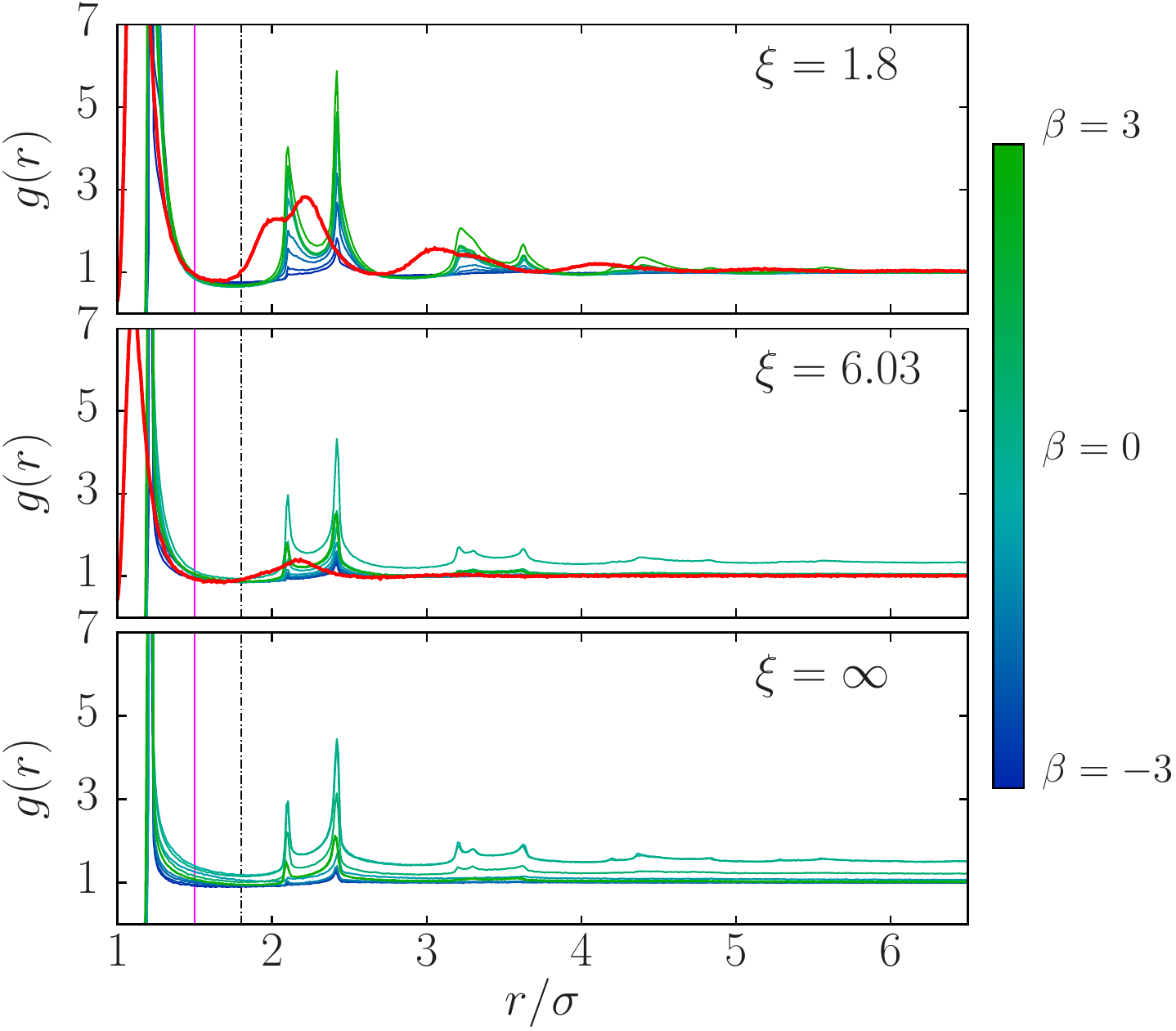}  
\caption{Radial distribution functions $g(r)$ for three different values of $\xi=\left\lbrace 1.8, 6.03,\infty \right\rbrace$. The color code of the curves is as a function of $\beta$. The $g(r)$s for ABP are plotted in red. Black line represents the chosen value of $r=r_{cl}$ for squirmers, while pink line is $r_{cl}$ for ABP.} \label{gr}
\end{flushleft}
\end{figure}

Even though  the $g(r)$ depends on the hydrodynamic feature and the interaction strength, 
in Figure \ref{gr} we show that a value of $r_{cl} = 1.8 \sigma$ is compatible with all chosen values of $\xi$ and $\beta$. 
Whereas the $g(r)$s for  ABP have their first minimum at $r_{cl} = 1.5 \sigma$, 
this difference between squirmers and ABP come from the fact that squirmers needs 
a soft repulsive potential for short distance ($\sim 1.1\sigma$) in order to avoid overlapping among them.

One important remark about the radial distribution function around $r_{cl}$ is that it is quite flat, being the first and second coordination shell not too close to each other. Therefore, choosing any value within this minimum is not going to affect the cluster size. 

\begin{figure}[h!]
\begin{center}
\includegraphics[clip,scale=0.55]{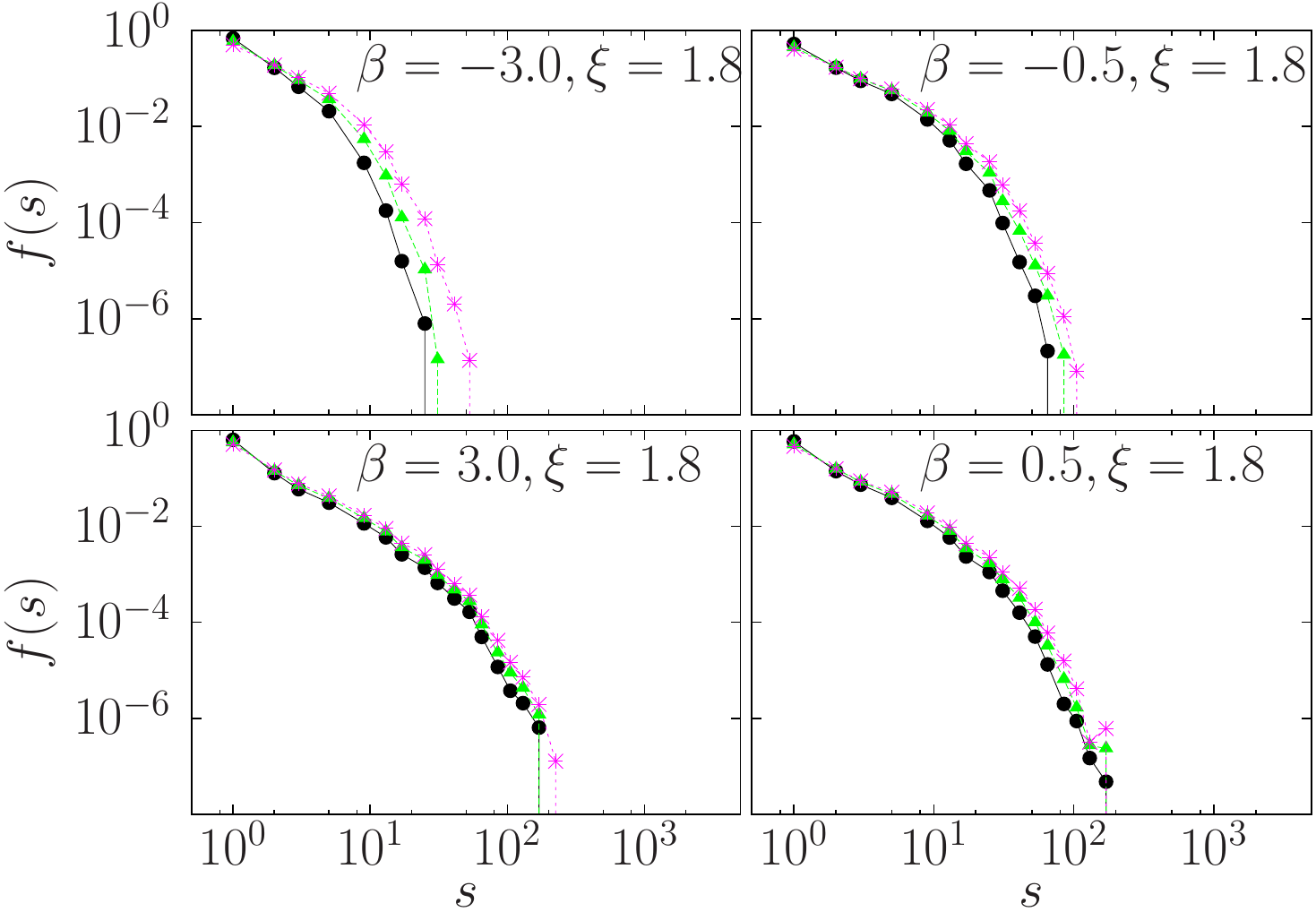}  
\caption{Cluster-size distribution $f(s)$ for values of $\xi=1.8$. Black circles correspond to $r_{cl}=1.3 \sigma$, green triangles to $r_{cl}=1.5 \sigma$, magenta stars to  $r_{cl}=1.8 \sigma$. The chosen values for $\beta$ and $\xi$ are indicated in the legend.} \label{fs_cutoff}
\end{center}
\end{figure}

In Figure \ref{fs_cutoff}, we show the cluster-size distribution functions for different values of $\beta$ at $\xi=1.8$, using three different values of $r_{cl}=\left\lbrace 1.3,1.5,1.8 \right\rbrace \sigma$. As expected, the CSDs follow the same power law with an exponential tail behaviour for the three values of $r_{cl}$, 
even though they move to larger clusters for larger values of $r_{cl}$ for pushers (that are less ordered, as shown in Figure \ref{gr}). 


\subsection{Values of $\Delta s_i$ used to compute  $f(s)$\label{sec:deltas}}

When computing the cluster-size distribution, we group clusters of an average size $s_i$ in bins with width $\Delta s_i=(s_{i,max}-s_{i,min})$ as in Ref. \cite{chantalbins2012} (Table~\ref{intervals}).

\begin{table}[h!]
\begin{center}
\caption{Cluster-size intervals $\Delta s_i=[s_{i,max},s_{i,min}]$, centred around $s_i$.}
\label{intervals}
\begin{tabular}{lcc|cc|cc}\hline\hline                         &
$[s_{i,max},s_{i,min}]$ & $s$ & $[s_{i,max},s_{i,min}]$ & $s$ & $[s_{i,max},s_{i,min}]$ & $s$      \\ \hline
$\mbox{}$ & 1& $ 1        $ & [36,45]& $ 41 $ & [451,600]&$526$    \\
$\mbox{}$ & 2& $ 2        $ & [46,55]& $ 51 $ & [601,1000] &$801$   \\
$\mbox{}$ & 3& $ 3        $ & [76,95]& $ 81 $ & [1001,1300] &$1151$   \\
$\mbox{}$ & [4,7]& $ 5    $ & [96,120]& $ 109  $ & [1301,2000] &$1651$   \\
$\mbox{}$ & [8,11]& $ 9   $ & [121,140]& $ 131 $ & [2001,3500] &$2751$   \\
$\mbox{}$ & [12,15]& $ 14 $ & [141,200]& $ 171 $ & [3501,5000]& $4251$   \\
$\mbox{}$ & [16,21]& $ 17 $ & [201,250]&$ 226 $ & [5001,7000]&$6001$   \\ 
$\mbox{}$ & [22,27]& $ 25   $ & [251,300]& $ 276 $ & [7001,10000] &$8501$   \\
$\mbox{}$ & [28,35]& $ 32 $ & [301,450]& $ 376 $ &    \\						 
\hline\hline
\end{tabular}
\end{center}
\end{table}

\subsection{Finite-size effects} \label{finiteffect}

\begin{figure}[h!]
\begin{center}
\includegraphics[clip,scale=0.6]{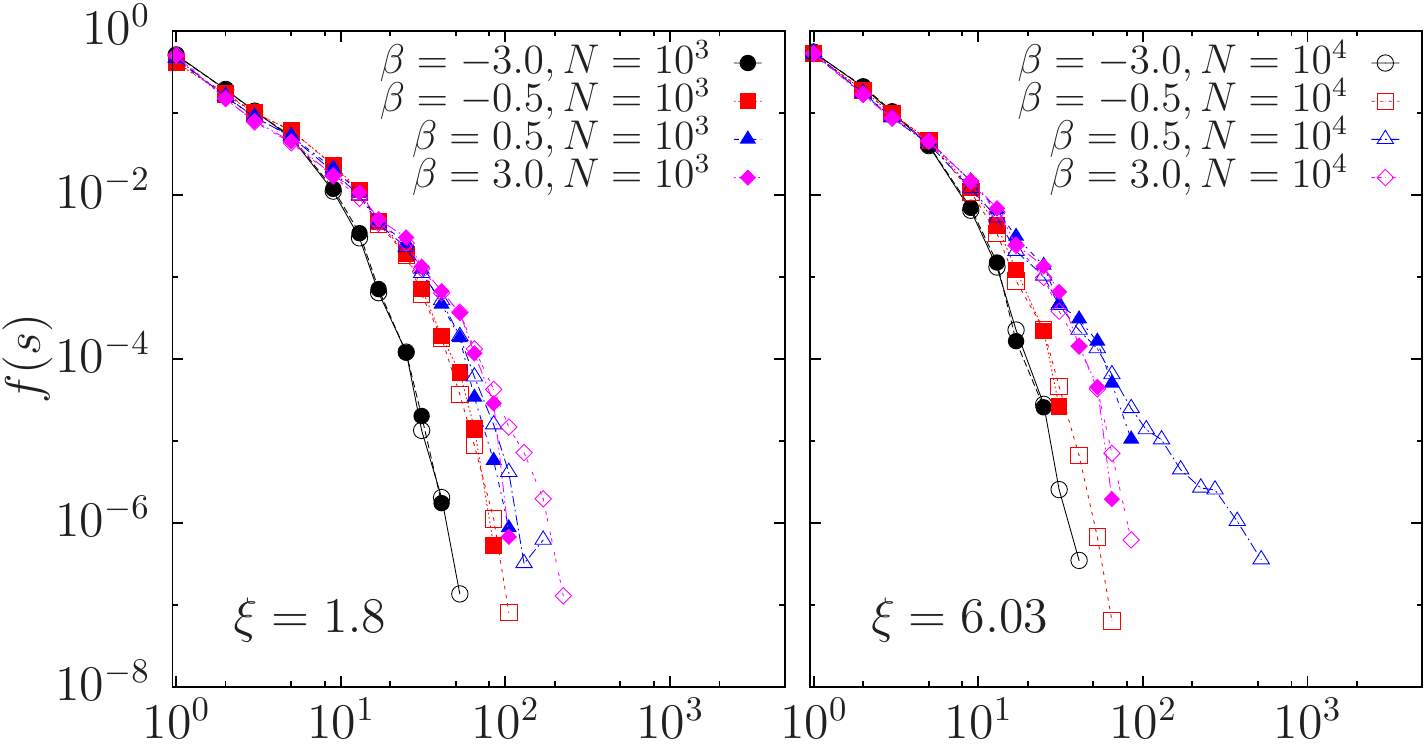} \\
 \includegraphics[clip,scale=0.725]{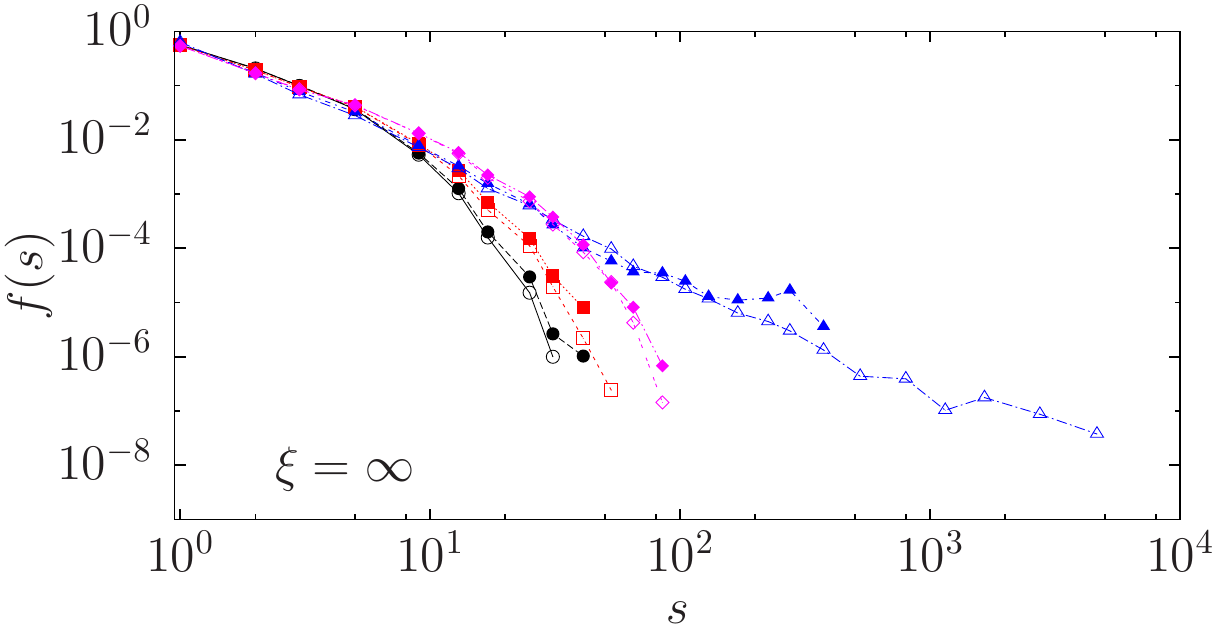} 
\caption{Cluster-size distribution $f(s)$ for values of $\xi=\left\lbrace 1.8, 6.03,\infty \right\rbrace$. In all panels, solid symbols are for small systems with $N=10^3$ whereas solid symbols are for big systems with $N=10^4$.}
\label{f_s_bp0_5}
\end{center}
\end{figure}

	In Figure~\ref{f_s_bp0_5} we represent the cluster-size distribution for different values of  $\xi$ and $\beta$ evaluated at the same $\phi=0.1$ for  two different system sizes, $N=10^3$ with filled symbols and $N=10^4$ with empty symbols. 

Finite-size effects are not present neither when $\xi=1.8$ nor when $\xi=6.03$: in both cases, $f(s)$ obtained for the small system size coincides with the one obtained for the larger system. However, when $\xi=\infty$ the results obtained for the small and large system only coincide for small clusters and deviate for larger ones. This is due to  the fact that at $\xi=\infty$ cluster formation is originated by hydrodynamics (long range) interactions, thus affected by the system size.

\begin{figure}[h!]
\begin{center}
\includegraphics[clip,scale=0.7]{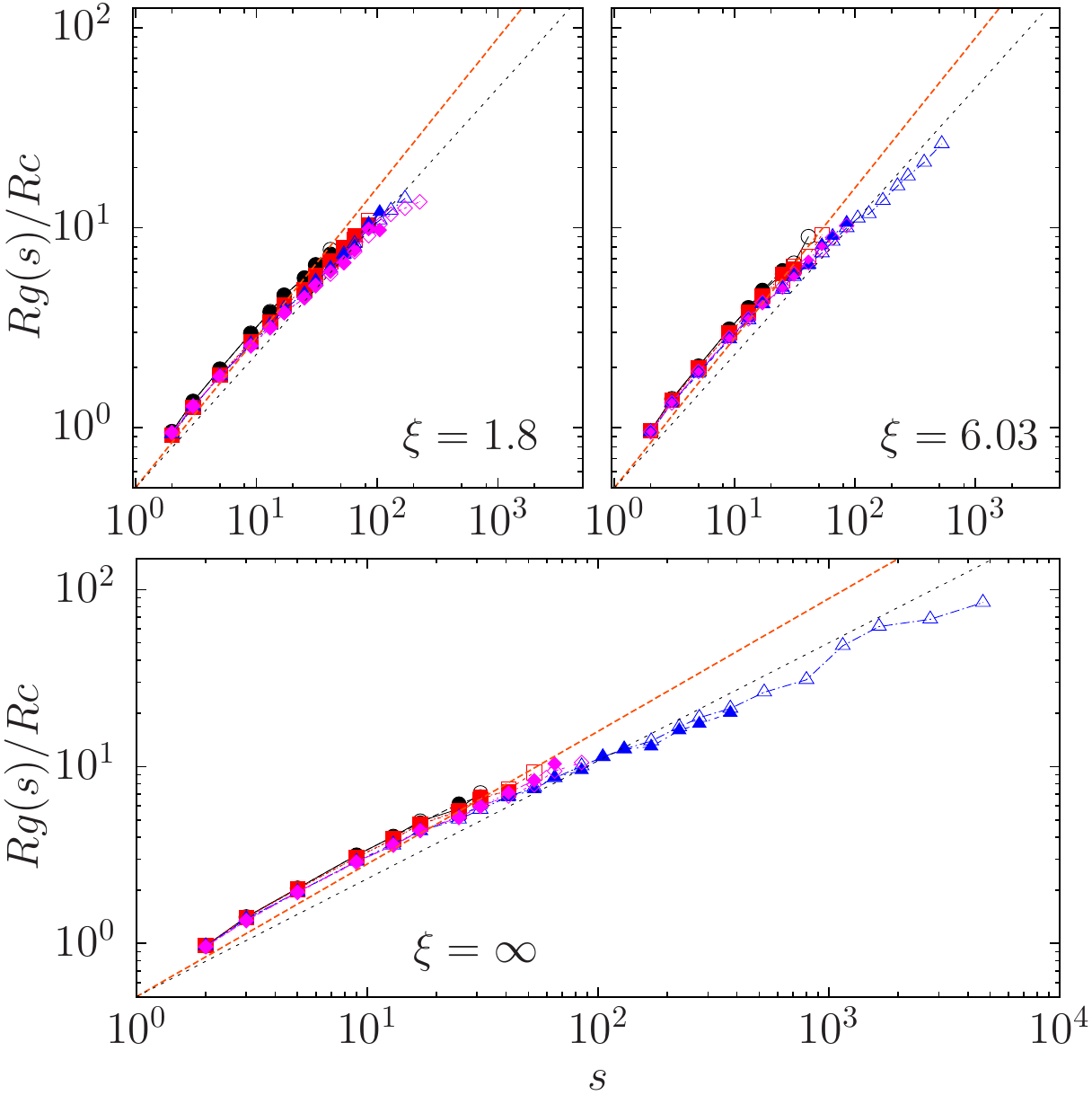}  
\caption{Radius of gyration $R_g(s)$ (normalized by the particle's radius $R_c$) 
as a function of the cluster-size for values of $\xi=\left\lbrace 1.8, 6.03,\infty \right\rbrace$. The symbols are the same as in Figure \ref{f_s_bp0_5}. Orange dashed curve and pointed black curve are a guide to the eye and represent $s^{3/4}$ and $s^{2/3}$ respectively.}
\label{RG_bp05}
\end{center}
\end{figure}

We now evaluate the clusters' radius of gyration for two different system sizes. In Figure~\ref{RG_bp05} we represent the radius of gyration normalized by the particle's radius 
for different values of  $\xi$ and $\beta$. The symbols are the same as in Figure \ref{f_s_bp0_5}.

As shown in the figure, there are not finite-size effects in the radius of gyration of the clusters despite the chosen value of $\xi$.

\end{document}